\documentclass[a4paper,11pt]{article}
\pdfoutput=1
\usepackage{amsmath}
\usepackage{graphicx}
\usepackage{amsfonts}
\usepackage{graphicx}
\usepackage{enumerate}
\DeclareFontFamily{U}{rsf}{}
\DeclareFontShape{U}{rsf}{m}{n}{
  <5> <6> rsfs5 <7> <8> <9> rsfs7 <10-> rsfs10}{}
\DeclareMathAlphabet\Scr{U}{rsf}{m}{n}

\newcommand{\be}{\begin{equation}}
\newcommand{\ee}{\end{equation}}
\newcommand{\ba}{\begin{eqnarray}}
\newcommand{\ea}{\end{eqnarray}}
\mathchardef\varGamma="0100
\mathchardef\varDelta="0101
\mathchardef\varTheta="0102
\mathchardef\varLambda="0103
\mathchardef\varXi="0104
\mathchardef\varPi="0105
\mathchardef\varSigma="0106
\mathchardef\varUpsilon="0107
\mathchardef\varPhi="0108
\mathchardef\varPsi="0109
\mathchardef\varOmega="010A

\author{
  \begin{minipage}{0.97\linewidth}
   \vspace{-8cm} \begin{flushright}
CPHT-RR032.0513\\
ROM2F/2013/03
\end{flushright}
    \vspace{5cm}
    \vspace{1cm}
    \begin{center}
      \begin{normalsize}
        \textbf{Cezar Condeescu}$^{a,b}$, \textbf{Emilian Dudas}$^{c}$
     \end{normalsize}
    \end{center}
    \vspace{.3cm} \hspace{2.0cm}
    \begin{minipage}{1.75\linewidth}
      { \begin{footnotesize}
          \begin{itemize}
          \item[${}^a$] Sezione INFN e Dipartimento di Fisica\\
           Universit\`a di Roma ``Tor Vergata"\\
          Via della Ricerca Scientifica 1, 00133 Roma, Italy
          \item[${}^b$] Department of Theoretical Physics\\
          Horia Hulubei National Institute of Physics and Nuclear Engineering - IFIN-HH\\
          P.O. Box MG-6, M\u{a}gurele - Bucharest, 077125, Rom\^ania
          \item[${}^c$] Centre de Physique Th\'eorique, Ecole Polytechnique and CNRS\\
          91128 Palaiseau, France
          \end{itemize}
        \end{footnotesize}}
        {\footnotesize E-mail addresses: cezar.condeescu@roma2.infn.it, emilian.dudas@cpht.polytechnique.fr.}
    \end{minipage}
    \vspace{0.5cm}
  \end{minipage}
}

\date{}
\title{\vspace{1mm}
  \begin{huge}
    \textbf{Kasner solutions, climbing scalars and big-bang singularity}
  \end{huge}
}

\begin{document}
\begin{titlepage}
\maketitle
\thispagestyle{empty}

\begin{center}
    \textsc{Abstract}
  \end{center}

\noindent
We elaborate on a recently discovered phenomenon where a scalar field close to big-bang  is forced to climb a steep potential by its dynamics. We analyze the phenomenon in  more general terms by writing the leading order equations of motion near the singularity.
    We formulate the conditions for climbing to exist in the case of several scalars and after inclusion of higher-derivative corrections and we apply our results to some models of moduli stabilization. We analyze an example with steep stabilizing potential and notice again a related critical behavior: for a potential steepness above a critical value, going backwards towards big-bang, the scalar undergoes wilder oscillations, with the steep potential pushing it back at every passage and not allowing the scalar to escape to infinity.

Whereas it was pointed out earlier that there are possible implications of the climbing phase to CMB,  we point out here another potential application, to the issue of initial   conditions in inflation.

%\vspace*{-15cm}

\end{titlepage}

%\begin{flushright}
%{\today}
%\end{flushright}

\tableofcontents

\section{Introduction}

A generic feature of string theory solutions with supersymmetry breaking at the string scale is the presence of scalar fields obeying an exponential potential law.  This can have important consequences for cosmology, especially in the context of inflationary models \cite{inflation1}. Based on the earlier work \cite{halliwell,dm,townsend}, it was found in \cite{dks} that for steep enough potentials, scalar fields
are forced to climb their potential right after big-bang. Indeed, for a system consisting of an exponential potential for a scalar field minimally coupled to a (purely) time dependent homogeneous gravitational background, one can find the exact solutions to the field equations. The results show that, when the logarithmic slope of the potential exceeds a critical value, the solution where the scalar descends the scalar potential after big-bang disappears. The scalar field is forced to climb the potential for a finite amount of time after which it will start descending. A more refined scenario involving a scalar potential given by a sum of two exponentials can naturally lead to inflation after the climbing phase. The consequences for the CMB power spectrum in this type of scenario have been investigated in \cite{dkps}. An interesting fact from the string theory point of view is that the critical value is precisely realized in the simplest orientifold models\footnote{For earlier papers and reviews on orientifold models, see e.g.
\cite{orientifolds}.} with supersymmetry breaking at the string scale: tachyon-free non-BPS vacua \cite{bsb} and the KKLT scenario of moduli stabilization \cite{kklt}. A natural question that arises is how often a climbing scalar aries in string effective actions. An approach based on exact solutions is clearly limited to specific effective actions.

The analysis that we carry on in this paper is two fold. On the one hand we look for climbing behavior in more complicated scenarios involving multiple scalars or more generic gravitational backgrounds. A crucial fact about the climbing phenomenon is that the scalar starts to climb asymptotically close to the big-bang. This opens up the possibility to find algebraic criteria for the existence of a climbing scalar by solving the field equations in the leading order close to the big-bang cosmological singularity \footnote{In principle our analysis should be valid for any naked space-time singularity, though according to the Cosmic Censorship Hypothesis (CCH) the big-bang is the only one of this type \cite{Penrose:1999vj}. For observational consequences of other naked singularities, see \cite{Virbhadra:2002ju}. We thank K.S. Virbhadra for pointing out these references to us.}.
The typical solutions that we find are of Kasner type both for the metric and for the scalar fields, which in the case of homogeneous and isotropic universes reduce to FRW-like time evolution for the scale factor. For the case analyzed in \cite{dks} we recover the critical exponent from the condition that the scalar potential and its first derivative are (fractionally) less singular than the kinetic terms in the equations of motion. The descending Kasner solution disappears due to the fact that for a supercritical exponent the scalar potential becomes too singular. It would be interesting to compare what we find by asymptotic methods with other exact solutions \cite{augusto}. There is also a large literature on black-hole solutions and exact solutions of gravity-scalar systems that, by analytic continuation,  could be used for finding exact cosmological solutions \cite{Gouteraux1}.

There are two immediate generalizations of the case above that we consider. One of them is to relax the $SO(d)$ symmetry condition of the gravitational background, with $d$ being the number of space dimensions (thus excluding time). It is easy to see that in this case one can always find asymptotic solutions where the scalar is allowed to descend immediately after big-bang. The other generalization that we consider is to have a model with an arbitrary number of scalars. We find exact solutions generalizing the Lucchin-Matarrese attractor \cite{lm} for the case of a multi-exponential potential. These solutions are characterized by the fact that the potential cannot be neglected asymptotically and that they require fine-tuned ``initial" conditions.
We give necessary and sufficient conditions for having climbing in a model with two scalar fields and an exponential potential. We also provide a simple sufficient condition for a model with an arbitrary number of scalar fields. Furthermore, we analyze with the same method some string effective actions with moduli stabilization. In the KKLT toy model there is always a descending scalar solution but with fine-tuned ``initial" condition. For generic ``initial" conditions there is a climbing scalar. We also show that the addition of the axion-dilaton with linear superpotential will spoil the climbing in the KKLT scenario.\\
There is an intriguing example consisting of one scalar field with a stabilizing potential given by a sum of two exponential terms with supercritical exponents. We were not able to find in this case any Kasner solution close to the big-bang. In addition to the obvious de Sitter solution with the scalar sitting at the minimum of the potential, there are solutions with wild oscillations
of the scalar field near big-bang. More precisely, going backwards in time towards big-bang,
the scalar undergoes wilder and wilder oscillations, with the steep potential sending the scalar
back at each passage. The scalar has therefore no well-defined limit at the singularity, which is otherwise reached in finite time. In counter-distinction, for subcritical potentials going backwards in time, the oscillations of the scalar field are amplified until a Planckian time, before which the scalar escapes and goes to infinity at the singularity.

On the other hand we consider the first $\alpha'$ corrections to the model with one scalar field. We start with an action containing quadratic terms in the curvature and (non-minimal) couplings of the scalar field to the curvature tensor and scalar and to the Ricci tensor with arbitrary coefficients. We impose first the condition that the action is ghost free, that is the equations of motion do not contain derivatives higher than two (for an arbitrary gravitational background). Afterwards we ask that the ghost free action preserve the Kasner solution that we find at zeroth order. It turns out that there is a unique (up to normalization) action at quadratic level in the curvature that is both ghost-free and that preserves the Kasner solution and hence the original climbing behavior. It is far fetched to assume that this exact combination can be realised in string theory. Even so, there is an infinite series of $\alpha'$ corrections that can destroy the climbing. From this point of view, to make a full analysis, it would be necessary to find a string model with a climbing scalar in the effective theory and with an exact CFT description available. We solve asymptotically two other examples, one based on DBI type correction $\sqrt{1+(\nabla \Phi)^2}$ and one based on the ghost free operator $G^{\mu\nu} \nabla_\mu\Phi \nabla_\nu\Phi$, where the addition of the extra higher order term(s) to the original action changes the solution in such a way that the climbing disappears.

The paper is organized as follows. In Section 2 we review the model considered in \cite{dks} with one scalar field from the point of view of an asymptotic (local) analysis in the vicinity of the big-bang. We reproduce the same conclusions and we further show that, when relaxing the $SO(d)$ symmetry condition on the gravitational background metric, one can always find descending solutions. Section 3 considers solutions for which the potential cannot be neglected asymptotically for systems with $n$ scalar fields. Section 4 is dedicated to a model with wild
scalar oscillations at the  big-bang singularity. This arises because the potential considered is a steep stabilizing potential, more precisely a sum of two exponentials with critical (or supercritical) exponents. In this case, both the climbing and descending Kasner solutions disappear as they lead to a too singular potential. We find a de Sitter solution with the scalar field sitting at the minimum of the potential. On a toy model with critical exponent we argue that the period of oscillations of the scalar field becomes smaller and smaller approaching the big-bang. This forces the scalar to oscillate forever going backwards in time towards the singularity, and therefore it has no well defined limit. In Section 5 we consider the effect of higher derivative corrections or $\alpha'$ corrections to the model with one scalar field exhibiting climbing. Considering the curvature square corrections and the couplings to the scalar field we find that there is a unique action, up to a normalization factor, which contains only up to second order derivatives in the field equations and which preserves the climbing Kasner solution for the scalar field. Section 6 is dedicated to models with multiple scalar fields. We give algebraic criteria on the initial data, e.g. exponents of the scalar potential, which imply the existence of a climbing scalar. Particular attention is given to the case of two scalar fields where we find necessary and sufficient conditions for climbing. In the general case we give only a sufficient condition. An exhaustive analysis is possible for the case of several scalars, but it is beyond the scope of this paper. Finally, in Section 7 we apply the results from Section 6 to string models with moduli stabilization. In particular we (re)consider the KKLT scenario and analyze the effect of adding to the relevant dynamics the axion-dilaton field subject to a polynomial superpotential. We show that in the case of a linear superpotential the phenomenon of climbing disappears. Two appendices collect some details about the higher derivative corrections and about how to transform quantities from the Einstein frame to the string frame for the effective action considered in Section 5.

%%%%%%%%%%%%%%%%%%%%%%%%%%%%%%%%%%%%%%%%%%%%%%%%%%%%%%%%%%%%%%%%%%%%%%%%%%%%%%%%%%%%%%%%%%%%%%%%%%%%%%%%%%%%%%%%%%%%%%%%%%%%%%%%%
\section{ Kasner solutions and climbing: one scalar field}

We reconsider in this section, from a different point of view, the simplest model with a climbing scalar \cite{dks}. It consists of a single scalar field minimally coupled to gravity with an exponential potential. The metric background is chosen to depend only on time and to be $SO(d)$ symmetric with $d$ being the number of space dimensions. The conclusion of \cite{dks}, based on finding the exact solutions of the field equations, was that the scalar is forced to climb the potential right after big-bang if the logarithmic slope of the potential is above a certain critical value $\lambda_c$. We shall reproduce the same result by means of an asymptotic analysis
near the big-bang singularity. Throughout the paper, we use the terminology of Kasner solutions
even for solutions with maximal space symmetry, which should more appropriately be called FRW. The reason is the emergence of Kasner spheres, like in general relativity with asymmetric scale factors, similar as in our eq. (\ref{soln}), but with the Kasner vector having as components the velocity of various scalar fields, like for example in (\ref{e6}) and (\ref{e12}) later on.

%%%%%%%%%%%%%%%%%%%%%%%%%%%%%%%%%%%%%%%%%%%%%%%%%%%%%%%%%%%%%%%%%%%%%%%%%%%%%%%%%%%%%%%%%%%%%%%%%%%%%%%%%%%%%%%%%
\subsection{Kasner solutions with $SO(d)$ Symmetry}

 A scalar field with potential $V$ propagating in a gravitational background is described by the following action\footnote{We use the ``mostly plus" metric throughout the paper.}
\begin{equation}
S_0 \ = \ \frac{1}{2}\int d^{d+1}x \sqrt{-g}\left[R-\frac{1}{2}(\nabla\Phi)^2-V(\Phi)\right] \ .
\label{action}
\end{equation}
We leave the potential $V(\Phi)$ unspecified for the moment, in order to describe the general strategy for finding asymptotic solutions close to the big-bang singularity. We start from the beginning with a gravitational background with $SO(d)$ symmetry of the following form
\begin{equation}
ds^2 \ = \ -  dt^2 \ + \ e^{2 A(t)} \ \sum_{i=1}^d(dx^i)^2 \ . \label{sod1}
\end{equation}
As we will see later on, relaxing the $SO(d)$ symmetry condition generically allows to find
always descending solutions.
The equations of motion in the aforementioned background are readily found to be
\begin{equation}
\begin{split}
&\frac{d(d-1)}{2}\dot A^2\,=\,\frac{1}{4}\dot\Phi^2+\frac{1}{2}V(\Phi) \ , \\
&\frac{d(d-1)}{2}\dot A^2+(d-1)\ddot A\,=\,-\frac{1}{4}\dot\Phi^2+\frac{1}{2}V(\Phi) \ , \\
&\ddot\Phi+d\dot A\dot\Phi\,=\,-\frac{\partial V}{\partial \Phi} \ . \\
\end{split} \label{sod2}
\end{equation}
We search for asymptotic solutions valid in a small neighbourhood of the Big Bang time $t_0$. We assume that the terms involving the scalar potential $V$ in the equations of motion (\ref{sod2}) can be neglected in the limit $t\rightarrow t_0$.
Neglecting the scalar potential then yields the following solutions for scale factor $A(t)$ and for the scalar field $\Phi(t)$
\begin{equation}
A \ = \ \frac{1}{d}  \ \ln(t-t_0) + A_0 \qquad , \qquad \Phi\ =\ \pm \sqrt{\frac{2(d-1)}{d}}
\ln (t-t_0) + \Phi_0 \ . \label{sod3}
\end{equation}
The integration constants $A_0$ and $\Phi_0$ fix the value of the fields at a reference time after the Big Bang, while $t_0$ specifies the Big Bang time. Without loss of generality we can choose
\begin{equation}
A_0 \ = \ 0 \qquad , \qquad t_0 \ = \ 0 \ ,
\end{equation}
so that the asymptotic solutions are of the form
\begin{equation}
A \ = \ b \ \ln t \qquad , \qquad \Phi = \ p\  \ln t +\Phi_0 \ , \label{sod03}
\end{equation}
with the constants $a$ and $p$ having the values
\begin{equation}
b \ = \ \frac{1}{d} \qquad , \qquad p \ = \ \pm \sqrt{\frac{2(d-1)}{d}} \ . \label{sod8}
\end{equation}
We keep the ``initial" condition $\Phi_0$ since, as we will see later, it plays an important role in finding exact attractor solutions for late times.
Notice that there are two solutions for the scalar field $\Phi$, one with positive velocity and one with negative velocity. Suppose that the potential $V(\Phi)$ is a monotonically increasing function. Then the solution with positive velocity describes a climbing scalar, whereas the solution with negative velocity describes a descending scalar.

With the solutions in eq. (\ref{sod3}) it is easy to see that the asymptotic behavior of $R$ and $(\nabla \Phi)^2$ is the following
\begin{equation}
R\, \sim\,  t^{-2} \ , \qquad  \ (\nabla\Phi)^2\, \sim\, t^{-2} \ .
\end{equation}
The same behavior is found for the various terms they generate in the equations of motion. In view of this it follows that the solutions we found in eq. (\ref{sod3}) are valid as long as they satisfy the following consistency constraint
\begin{equation}
V(\Phi(t)) \ , \ \frac{\partial V(\Phi(t))}{\partial \Phi} \ \sim \ O(t^{-2+\epsilon}) \ , \label{sod5}
\end{equation}
that is, they are fractionally negligible with respect to the kinetic part. In other words, any free solution to the equations of motion can describe the asymptotic behavior of the system if and only if the constraint above is satisfied.
Let us consider the case, examined in \cite{dks}, of an exponential scalar potential
\begin{equation}
V(\Phi) \ = \ 2 \ \alpha \ e^{\lambda\Phi} \ , \label{exp}
\end{equation}
with $\alpha$ and $\lambda$ positive real constants\footnote{Our conventions in this paper differ
from the ones in \cite{dks} and \cite{dkps}. The connection between our exponent $\lambda$ and
$\gamma$ from \cite{dks,dkps} is $\lambda = \sqrt{\frac{2d}{d-1}} \gamma$. Our scalar field is
related to the field $\varphi$ of these references via $\Phi = \sqrt{\frac{2(d-1)}{d}} \varphi$.}, such that the potential is an increasing function of $\Phi$ and also bounded from below. Notice that the replacement $\lambda \rightarrow -\lambda$ would only exchange the climbing with the descending solution and viceversa. Accordingly, it can be eliminated by redefining the scalar field $\Phi\rightarrow -\Phi$. The condition in eq. (\ref{sod5}) then implies that the asymptotic solutions in eq. (\ref{sod3}) are valid as long as the following inequality is satisfied
\begin{equation}
\lambda \ p \ > \ - \ 2 \ . \label{sod6}
\end{equation}
The solution describing a climbing scalar ($p>0$), satisfies automatically this condition for any given positive constant $\lambda$. However, the descending solution ($p<0$) exists if and only if $\lambda$ satisfies the following inequality
\begin{equation}
\lambda\ < \ \sqrt{\frac{2d}{d-1}}\equiv \lambda_c \ .
 \ \label{sod9}
\end{equation}
{\bf Conclusion}: If $\lambda \geq \lambda_c$ then only the climbing solution exists.
This is indeed what has been found in \cite{dks} by a direct analysis of the exact equations of motion using a parametric time $\tau$ related to the cosmological time via
$dt \sim d \tau / \sqrt{V} $.

A useful viewpoint is to reverse the arrow of time and go backwards
towards the big-bang. In the subcritical case $\lambda < \lambda_c$, starting at a given time with an initial velocity such that the scalar climbs the potential, there is always a minimum velocity such that the kinetic term is winning over the potential terms and the scalar  continues to climb until the singularity is reached for infinite field value. In the critical and supercritical case
$\lambda \ge \lambda_c$, the potential becomes too abrupt and, irrespective of how
fast the scalar is smashed again the potential wall, it will stop at some point and revert its motion before reaching the singularity. This is nothing but the time reversal description of the climbing phenomenon discovered in \cite{dks}.

%%%%%%%%%%%%%%%%%%%%%%%%%%%%%%%%%%%%%%%%%%%%%%%%%%%%%%%%%%%%%%%%%%%%%%%%%%%%%%%%%%%%%%%%%%%%%%%%%%%%%%%%%%%%%%%%%%%%%%%%%%%%%%%%%

%%%%%%%%%%%%%%%%%%%%%%%%%%%%%%%%%%%%%%%%%%%%%%%%%%%%%%%%%%%%%%%%%%%%%%%%%%%%%%%%%%%%%%%%%%%%%%%%%%%%%%%%%%%%%%%%%%%%%%%%%%%%%%%%
\subsection{Kasner solutions without symmetry: no climbing}
\label{generalkasner}

In the following we relax the $SO(d)$ symmetry condition of the gravitational background, thus allowing different scale factors depending on direction.
This corresponds to considering Kasner type metric tensors \cite{kasner}
\begin{equation}
ds^2 \ = \ - dt^2 + \sum_{i=1}^de^{2 A_i(t)}(dx^i)^2 \ ,
\label{kasner}
\end{equation}
with the functions $A_i(t)$ distinct of the form $A_i =b_i\ln t$. We search for asymptotic solutions to the equations of motion with metric given in eq. (\ref{kasner}). Taking into account the form of the gravitational background we have the following system of differential equations
\begin{equation}
\begin{split}
&\frac{1}{2}\left[\left(\sum_{i=1}^d\dot A_i\right)^2-\sum_{i=1}^d\dot A_i^2\right]\, =\, \frac{1}{4}\dot\Phi^2+\frac{1}{2}V(\Phi) \ , \\
&\sum_{i\neq k}\ddot A_i+\frac{1}{2}\left[\left(\sum_{i=1}^d\dot A_i\right)^2+\sum_{i=1}^d\dot A_i^2\right]-\dot A_k\sum_{i=1}^d\dot A_i \, =\, -\frac{1}{4}\dot\Phi^2+\frac{1}{2}V(\Phi) \qquad {\rm\ for\ } k=1,...,d \ , \\
&\ddot\Phi+\dot\Phi\sum_{i=1}^d\dot A_i\, =\, -\frac{\partial V}{\partial \Phi} \ . \\
\end{split}
\end{equation}
Similar as for the $SO(d)$ symmetric background, one can show that, for the case of an asymptotically negligible scalar potential (and first derivative), the solutions for the scale factors $A_i(t)$ and for the scalar field $\Phi$ have to be of the form
\begin{equation}
A_i \ = \ b_i \ln t \qquad , \qquad \Phi \ = \ p \ln t +\Phi_0\ ,
\end{equation}
in the limit $t\rightarrow 0$, that is close to the Big Bang.
Neglecting the terms arising from the scalar potential one is left with the following constraints to be satisfied by the parameters $b_i$ and $p$:
\begin{equation}
\begin{split}
&\frac{1}{2}\left[\left(\sum_{i=1}^d b_i\right)^2-\sum_{i=1}^db_i^2\right] \ = \ \frac{1}{4}p^2 \ , \\
&-\sum_{i\neq k}b_i +\frac{1}{2}\left[\left(\sum_{i=1}^d b_i\right)^2+\sum_{i=1}^db_i^2\right]-b_k\sum_{i=1}^db_i
\ = \ - \frac{1}{4}p^2 \ , \\
&-p+p\sum_{i=1^d}b_i \ = \ 0 \ . \\
\end{split}
\end{equation}
Furthermore, one can show that the algebraic constraints above reduce to
\begin{equation}
\sum_{i=1}^db_i \ = \ 1 \qquad , \qquad \sum_{i=1}^db_i^2+\frac{1}{2}p^2 \ = \ 1 \ .
\label{soln}
\end{equation}
In general relativity without scalar fields (equivalent to $p=0$ in the eqs. above), these
conditions tell us that $b_i$ live at the intersection between a Kasner sphere and a hyperplane,
as found in the original paper by  Kasner \cite{kasner}.
 For a given solution for the metric, that is coefficients $b_i$ satisfying $\sum_{i=1}^d b_i=1$, we have two solutions for the scalar field with parameter $p$ given by
\begin{equation}
p\, =\, \pm\sqrt{2\left(1-\sum_{i=1}^d b_i^2\right)} \ .
\end{equation}
Let us consider again the case of an exponential scalar potential, $V(\Phi) \sim e^{\lambda\Phi}$.
The descending solution exists as long as we satisfy $\lambda p > -2$. Notice that, in this case, one can always tune the parameters $b_i$ satisfying the first eq. in (\ref{soln}) such that the resulting solution for $p$ can be made arbitrarily small. Indeed, take for example
\begin{equation}
b_1=1-\epsilon \qquad , \qquad b_2=...=b_d=\frac{\epsilon}{d-1}
\end{equation}
were $\epsilon>0$ can be taken arbitrarily small. Then it is easy to see that we have
\begin{equation}
p^2\sim \epsilon
\end{equation}

As a consequence, for a given finite $\lambda$ one can always find a descending solution such that $\lambda p >-2$. Hence the climbing phenomenon disappears if one relaxes the $SO(d)$ symmetry considered in the previous section.

%%%%%%%%%%%%%%%%%%%%%%%%%%%%%%%%%%%%%%%%%%%%%%%%%%%%%%%%%%%%%%%%%%%%%%%%%%%%%%%%%%%%%%%%%%%%
\section{Exact and mixed-type solutions for $n$ scalar fields}

In this section we relax the condition of having negligible potential close to big-bang and allow
for the possibility that the potential and kinetic contributions are similar. Some of the solutions below are generalizations of the original Lucchin-Matarrese exact solution \cite{lm},
which has in turn the interpretation of late-time attractor for the more general solution
described in \cite{dks}, \cite{dkps}.

Let us start with the case of one scalar field.  We search for solutions such that the scalar potential cannot be neglected for $t\rightarrow 0$.
Suppose that we have a solution for the scalar field $\Phi$ such that the potential behaves as
\begin{equation}
V(\Phi(t)) \sim t^{-2} \ . \label{e1}
\end{equation}
Then for the exponential scalar potential we have that the solution for $\Phi$ must be of the form
\begin{equation}
\Phi = -\frac{2}{\lambda} \ln t +\Phi_0 \ , \label{e2}
\end{equation}
or, in other words, we have descending scalar with the constant $p$ now satisfying $\lambda \ p = -2$.

 We search for solutions of the form $A=b\ln(t)$, since we must have $\dot A ^2\sim t^{-2}$. All terms are of the order $t^{-2}$ and one is reduced to the following algebraic system
\begin{equation}
\begin{split}
&\frac{d(d-1)}{2\,} b^2 \ = \ \frac{p^2}{4}+ {\tilde \alpha} \ , \\
&\frac{d(d-1)}{2}\, b^2-(d-1)b \ = \ - \frac{p^2}{4}+ {\tilde \alpha} \ , \\
&-p+d\,b\,p \ = \ - 2 \lambda {\tilde \alpha} \ , \
\lambda p \ = \ - 2 \ . \\
\end{split} \label{sod10}
\end{equation}
The system above has a solution if and only if the integration constant ${\tilde \alpha}
= \alpha\, e^{\lambda \Phi_0}$ satisfies the following equation
\begin{equation}
{\tilde \alpha} \ = \ \frac{\lambda_c^2 - \lambda^2}{\lambda^4} \ . \label{sod12}
\end{equation}
In this case, the constant parameters $b$ and $p$ determining the signs of the velocities of the scale factor and scalar field respectively,  are given by
\begin{equation}
b \ = \ \frac{1}{d} \frac{\lambda_c^2}{\lambda^2} \qquad , \qquad p  \ = \ - \frac{2}{\lambda} \ . \label{sod11}
\end{equation}
Since ${\tilde \alpha}$ is strictly positive, then making use of eq. (\ref{sod12}) it follows that we must have the condition
\begin{equation}
\lambda \ < \ \sqrt{\frac{2d}{d-1}} \equiv \lambda_c \ . \label{sod13}
\end{equation}
Notice that this descending solution is actually exact for all times and was found originally
by Lucchin and Matarrese \cite{lm}, that noticed that it can generate a power-law inflationary universe. Moreover,  eq. (\ref{sod12}) informs us that the integration constant $\Phi_0$ is exactly determined by the dynamics. It was shown in \cite{dks,dkps} that more general solutions all tend asymptotically for late time to this
Lucchin-Matarrese attractor.
Moreover, one can also show that no solutions exist if the exponential scalar potential is more singular than $t^{-2}$.

We consider now the generalization of such solution for a system of $n$ scalar fields. Einstein
and scalar fields eqs. are
\begin{align}
&\frac{d(d-1)}{2} \dot A^2 =\frac{1}{4}\sum_{i=1}^n\dot\Phi_i^2+\frac{1}{2}V(\Phi_1,...,\Phi_n)
\ , \nonumber \\
 &\frac{d(d-1)}{2} \dot A^2+(d-1)\ddot A=  -\frac{1}{4}\sum_{i=1}^n\dot\Phi_i^2+\frac{1}{2}V(\Phi_1,...,\Phi_n) \ , \nonumber \\
 &\ddot \Phi_k + d \dot A\dot\Phi_k = -\frac{\partial V(\Phi_1,...,\Phi_n)}{\partial \Phi_k}
\ . \label{e3}
\end{align}
The scalar potential is taken to be of the following form:
\begin{equation}
V(\Phi_1,...,\Phi_n) \ = \ \sum_{i=1}^n 2\alpha_i \ e^{\sum_{j=1}^n\lambda_{ij}\Phi_j} \ ,
\label{e4}
\end{equation}
where the simple form of sum of exponentials allows to discuss both exact solutions and the climbing behavior in simple terms.
We look for exact solutions of the system above of the following form:
\begin{align}
&A=b\ln t+ A_0 \ , \\
&\Phi_k=p_k\ln t + \Phi_{0k} \ . \label{e5}
\end{align}
Notice that irrespective of the form of the scalar potential, combining the Einstein eqs.
tell us that the scalar fields speeds sit on a Kasner-like sphere with radius determined
by the scale factor
\begin{equation}
\frac{1}{2} \sum_i p_i^2 \ = \ (d-1) \ b \ . \label{e6}
\end{equation}
The scalar potential's behavior as a function of time is given by:
\begin{equation}
V(\Phi_1(t),...,\Phi_n(t)=\sum_{i=1}^n 2\alpha_i \  e^{\sum_{j=1}^n\lambda_{ij}\Phi_{0j}}t^{\sum_{j=1}^n\lambda_{ij}p_j}\equiv \sum_{i=1}^n2\tilde \alpha_i \ t^{-2} \ , \label{e7}
\end{equation}
where we have defined $\tilde \alpha _i$
\begin{equation}
\tilde \alpha_i = \alpha_i \ e^{\sum_{j=1}^n\lambda_{ij}\Phi_{0j}} \label{e8}
\end{equation}
and we imposed the following constraints
\begin{equation}
\sum_{j=1}^n \lambda_{ij}p_j \ = \ - 2 \qquad \textrm{for all}\ i=1,...,n \ . \label{e9}
\end{equation}
Suppose, for simplicity that the matrix $\lambda = \lambda_{ij}$ has non-zero determinant. Then one can solve uniquely the constraints above for the (constant) velocities of the scalar fields
\begin{equation}
p_i=-2\sum_{j=1}^n(\lambda^{-1})_{ij} \ , \label{ps}
\end{equation}
that define one unique vector on the Kasner sphere, where the exact solution is pointing.
In order to satisfy the equations of motion one has to impose the following algebraic constraints on the ``initial" conditions $\Phi_{0k}$ of the scalar fields contained in the redefined scalar potential parameters $\tilde \alpha_i$:
\begin{align}
&\frac{d(d-1)}{2}b^2 = \sum_{i,j,k=1}^n(\lambda^{-1})_{ij}(\lambda^{-1})_{ik}+\sum_{i=1}^n \tilde \alpha_i \ , \nonumber \\
&(-1+b\,d)\sum_{j=1}^n (\lambda^{-1})_{kj}=\sum_{i=1}^n\tilde \alpha_i\lambda_{ik} \qquad , \qquad \forall\ k=1,...,n
\label{e10}
\end{align}
The algebraic system above consists of $n+1$ equations determining the unknowns $\tilde \alpha_i$ and $b$. The solution is given by
\begin{align}
&b=\frac{1}{d}\lambda_c^2\sum_{i,j,k=1}^n(\lambda^{-1})_{ij}(\lambda^{-1})_{ik} \ , \nonumber
\\ &\tilde \alpha_i=\left[\lambda_c^2\sum_{j,k,l=1}^n(\lambda^{-1})_{jk}(\lambda^{-1})_{jl}-1\right]\sum_{j,k=1}^n(\lambda^{-1})_{kj}(\lambda^{-1})_{ki}
\ , \label{a}
\end{align}
that fix all the integration constants $\Phi_{0k}$.
Notice that we have the following relation
\begin{align}
\sum_{i,j,k=1}^n(\lambda^{-1})_{ij}(\lambda^{-1})_{ik}= \sum_{i=1}^n \left(\sum_{j=1}^n (\lambda^{-1})_{ij}\right)^2=\frac{1}{4}\sum_{i=1}^np_i^2 \geq 0 \ ,
\label{e12}
\end{align}
thus implying that the scale factor velocity is indeed positive $b>0$.
Let us define the quantities $x_i$ appearing in eq. (\ref{a})
\begin{align}
x_i=\sum_{j,k=1}^n(\lambda^{-1})_{kj}(\lambda^{-1})_{ki}=-\frac{1}{2} \sum_{k=1}^n p_k(\lambda^{-1})_{ki} \ . \label{e13}
\end{align}
Then we can rewrite eq. (\ref{a}) as
\begin{align}
\tilde \alpha_i = \left(\lambda_c^2\sum_{j=1}^nx_j -1\right)x_i \ . \label{e14}
\end{align}
The exact solution found in eqs. (\ref{ps}) and (\ref{a}) exist as long as the constraints in eq. (\ref{a}) are satisfied. In particular for stability of the scalar potential one must also have
\begin{align}
\tilde \alpha_i > 0 \qquad \forall\ i=1,...,n \ . \label{e15}
\end{align}
Since we have that
\begin{align}
\sum_{i=1}^n x_i \ = \ \frac{1}{4}\sum_{i=1}^n p_i^2 > 0 \ , \label{e015}
\end{align}
then there exists an index $k\in\{1,...,n\}$ such that $x_k>0$. From the condition $\tilde \alpha_k >0$ it immediately follows that the following condition is necessary (though not sufficient)
\begin{align}
\lambda_c^2\sum_{j,k,l=1}^n(\lambda^{-1})_{jk}(\lambda^{-1})_{jl}>1 \ .
\label{not sufficient}
\end{align}
Moreover, one has to satisfy $x_i > 0$
\begin{align}
\sum_{j,k=1}^n(\lambda^{-1})_{kj}(\lambda^{-1})_{ki}>0 \qquad \forall i
\label{e16}
\end{align}
The condition for an accelerating Universe in our notation is $b > 1$, which leads to the condition
\begin{align}
\sum_{i,j,k=1}^n(\lambda^{-1})_{ij}(\lambda^{-1})_{ik} >  \frac{d}{\lambda_c^2} =
\frac{d-1}{2} \ .
\label{acc1}
\end{align}
%%%%%%%%%%%%%%%%%%%%%%%%%%%%%%%%%%%%%%%%%%%%%%%%%%%%%%%%%%%%%%%%%%%%%%%%%%%%%%%%
{\bf One Scalar Field}

By particularizing the formulas above for the case of one scalar field one obtains again the results in eqs. (\ref{sod12}), (\ref{sod11}).
Thus the solution for the scale factor $A$ and scalar field $\Phi$ can be written as
\begin{align}
&A \,=\,\frac{1}{d}\frac{\lambda_c^2}{\lambda^2} \ln t + A_0 \ , \nonumber \\
&\Phi \,=\,-\frac{2}{\lambda}\ln t + \frac{1}{\lambda}\ln \frac{1}{\alpha \lambda^2}\left(\frac{\lambda_c^2}{\lambda^2}-1\right) \ . \label{e18}
\end{align}
The condition in eq. (\ref{not sufficient}) is actually sufficient in the case of one scalar field and it becomes
\begin{equation}
\lambda < \lambda_c \ .
\end{equation}
We recover therefore the Lucchin-Matarrese solution \cite{lm} displayed at the beginning of this Section. The condition for an accelerating Universe is $b > 1$ leading to
$\lambda < \frac{\lambda_c}{\sqrt{d}}$.

%%%%%%%%%%%%%%%%%%%%%%%%%%%%%%%%%%%%%%%%%%%%%%%%%%%%%%%%%%%%%%%%%%%%%
{\flushleft \bf Two Scalar Fields}

For the case of two scalar fields it is useful to define the matrix of coefficients and its inverse
\begin{equation}
\lambda = \left(
  \begin{array}{cc}
    \lambda_{11} & \lambda_{12} \\
    \lambda_{21} & \lambda_{22} \\
  \end{array}
\right) \qquad  , \qquad \lambda^{-1}=\frac{1}{\det \lambda} \left(
                                                          \begin{array}{cc}
                                                            \lambda_{22} & -\lambda_{12} \\
                                                            -\lambda_{21} & \lambda_{11} \\
                                                          \end{array}
                                                        \right) \ .
\label{e19}
\end{equation}
Then it is easy to see that the equations for $n$ scalar fields found earlier in the section now reduce to
\begin{align}
p_1= \frac{2}{\det \lambda} (\lambda_{12}-\lambda_{22}) \qquad , \qquad p_2 = \frac{2}{\det \lambda}(\lambda_{21}-\lambda_{11}) \ , \label{e019}
\end{align}
\begin{align}
&x_1= \frac{1}{(\det \lambda)^2}(\lambda_{22}^2+\lambda_{21}^2- \lambda_{12}\lambda_{22}-\lambda_{11}\lambda_{21}) \ , \\
&x_2=\frac{1}{(\det \lambda)^2}(\lambda_{11}^2+\lambda_{12}^2- \lambda_{12}\lambda_{22}-\lambda_{11}\lambda_{21}) \ , \label{e20}
\end{align}
\begin{equation}
b=\frac{1}{d}\frac{\lambda_c^2}{(\det \lambda)^2}\left[(\lambda_{11}-\lambda_{21})^2+(\lambda_{22}-\lambda_{12})^2\right]
\ . \label{e21}
\end{equation}
It is useful to show explicitly that the solutions written above can describe only two descending scalars. Indeed, this is a consequence of the following proposition.\vspace {0.2 cm}\\
{\bf Proposition.}
Let $\lambda\in GL_2(\mathbb{R}^+)$ be an invertible $2\times 2$ matrix with positive entries satisfying the following constraints
\begin{align}
&x_1  , x_2 > 0 \ , \label{e22}
\end{align}
then the following inequalities must hold
\begin{align}
p_1, p_2 <0 \ . \label{e23}
\end{align}
{\bf Proof.} We prove the result by reductio ad absurdum. There are two cases to consider. First let us assume that both $p_i$'s are positive. Then one of the following must hold
\begin{align}
&\lambda_{12}>\lambda_{22} \quad , \quad \lambda_{21}>\lambda_{11} \quad , \quad
\textrm{if}\ \det \lambda  > 0 \quad \quad
&\textrm{or} \nonumber \\
&\lambda_{12}<\lambda_{22} \quad , \quad \lambda_{21}<\lambda_{11} \quad , \quad \textrm{if}\ \det \lambda  < 0 \ . \label{e23}
\end{align}
Using the fact that we must have $\lambda_{ij}>0$, the contradiction follows immediately by multiplying the inequalities
\begin{align}
&\lambda_{12}\lambda_{21}>\lambda_{11}\lambda_{22} \qquad \textrm{if}\ \det \lambda >0 \ ,
\nonumber \\
&\lambda_{12}\lambda_{21}<\lambda_{11}\lambda_{22} \qquad \textrm{if}\ \det \lambda <0 \ .
\label{e24}
\end{align}
Hence $p_1$ and $p_2$ cannot be both positive. The second case to consider is when one of the $p_i$'s is positive and the other negative. Then one of the two must hold
\begin{align}
& \lambda_{12}>\lambda_{22} \quad , \quad \lambda_{21} <\lambda_{11} \quad \quad
\textrm{or} \nonumber\\
& \lambda_{12}<\lambda_{22} \quad , \quad \lambda_{21} >\lambda_{11} \ . \label{e25}
\end{align}
It is easy to see that one must have one of the following cases
\begin{align}
& \lambda_{12}\lambda_{22}>\lambda_{22}^2 \quad , \quad \lambda_{21}^2 <\lambda_{11}\lambda_{21} \quad
\quad \textrm{or} \nonumber\\
& \lambda_{12}^2<\lambda_{22}\lambda_{12} \quad , \quad \lambda_{21}\lambda_{11} > \lambda_{11}^2
\ . \label{e26}
\end{align}
The contradiction follows as the inequalities above imply that we have
\begin{equation}
x_1 < 0 \qquad \qquad \textrm{or} \qquad \qquad x_2  < 0 \ . \label{e27}
\end{equation}
Hence the only possibility is that both scalars are descending. This completes the proof. \hfill $\Box$
This solution is the generalization of the Lucchin-Matarrese attractor for the two fields case
and it probably describes the late time attractor behavior for any solution.
%%%%%%%%%%%%%%%%%%%%%%%%%%%%%%%%%%%%%%%%%%%%%%%%%%%%%%%%%%%%%%%%%%%%%%%%%%%%%%%%%
{\flushleft \bf Mixed Case for Two Scalar Fields}

Let us now consider the mixed case in which one of the two potential terms behaves as $t^{-2}$,
whereas the other is negligible.
Suppose therefore that the following relations hold
\begin{align}
&\lambda_{11}p_1+\lambda_{12}p_2 \ = \ - 2 \ , \nonumber \\
&\lambda_{21}p_1+\lambda_{22}p_2 \ > \ -2 \ . \label{e28}
\end{align}
Then asymptotically the scalar potential terms in the equations of motion are given by
\begin{equation}
V(\Phi_1(t),\Phi_2(t))\simeq 2\tilde \alpha_1 \ t^{-2} \qquad , \qquad
\frac{\partial V(\Phi_1(t),\Phi_2(t))}{\partial \Phi_i} \simeq 2\tilde\alpha_1 \ \lambda_{1i}
\ t^{-2} \ , \label{e29}
\end{equation}
where, as before, we have absorbed the integration constants $\Phi_{01}$ and $\Phi_{02}$ into a redefined constant $\tilde \alpha_1$
\begin{align}
\tilde\alpha_1 = \alpha_1 \ e^{\lambda_{11}\Phi_{01}+\lambda_{12}\Phi_{02}} \ . \label{e30}
\end{align}
One can find solutions to the equations of motion by solving the following system of algebraic equations
\begin{align}
&\frac{d(d-1)}{2}b^2=\frac{1}{4}\left(p_1^2+p_2^2\right)+\tilde \alpha_1 \ , \nonumber \\
&-p_1+d\,b\,p_1=-2\tilde\alpha_1 \ \lambda_{11} \ , \nonumber \\
&-p_2+d\,b\,p_2=-2\tilde\alpha_1 \ \lambda_{12} \ , \nonumber \\
&\lambda_{11}p_1+\lambda_{12}p_2 = - 2 \ . \label{e31}
\end{align}
The solution is then found to be
\begin{align}
&p_1=-\frac{2\lambda_{11}}{\lambda_{11}^2+\lambda_{12}^2} \qquad , \qquad p_2=-\frac{2\lambda_{12}}{\lambda_{11}^2+\lambda_{12}^2} \ , \nonumber \\
&b=\frac{1}{d} \frac{\lambda_c^2}{\lambda_{11}^2+\lambda_{12}^2} \qquad , \qquad \tilde \alpha_1 = \frac{\lambda_c^2-\lambda_{11}^2-\lambda_{12}^2}{(\lambda_{11}^2+\lambda_{12}^2)^2} \ ,
\label{e32}
\end{align}
describing again two descending scalars. The solution exists as long as we have the condition $\tilde \alpha_1 >0$ satisfied. Using the expression of $\tilde \alpha_1$ found above, on gets the following condition
\begin{align}
\lambda_{11}^2+\lambda_{12}^2 < \lambda_c^2 \ . \label{e33}
\end{align}
Furthermore, an accelerating Universe arises if $b > 1$, thus leading to the condition
\begin{align}
\lambda_{11}^2+\lambda_{12}^2 < \frac{\lambda_c^2}{d} \ . \label{e033}
\end{align}
Notice that these solutions are not exact as we have neglected one term in the scalar potential. Unlike the Lucchin-Matarrese type solutions found before these cannot be attractor solutions for late times. By construction they are valid only in the vicinity of the big-bang.
%%%%%%%%%%%%%%%%%%%%%%%%%%%%%%%%%%%%%%%%%%%%%%%%%%%%%%%%%%%%%%%%%%%%%%%%%%%%%%%%%%%%%%
{\flushleft \bf Two Scalar Fields with $ \det \lambda = 0$}

In this particular case, we search for solutions satisfying the following constraints
\begin{align}
&\lambda_{11}p_1+\lambda_{12}p_2=-2 \qquad , \qquad
\lambda_{21}p_1+\lambda_{22}p_2=-2 \ , \nonumber \\
&\det \lambda = \lambda_{11}\lambda_{22}-\lambda_{12}\lambda_{21}=0 \ . \label{e34}
\end{align}
It is easy to see that in order to have the equalities above satisfied the following relations must hold
\begin{align}
\lambda = \left(
           \begin{array}{cc}
             \lambda_{11} & \lambda_{22} \\
             \lambda_{11} & \lambda_{22} \\
           \end{array}
         \right) \qquad , \qquad
 \lambda_{11} p_1+\lambda_{22} p_2 = - 2 \ . \label{e35}
\end{align}
The scalar potential of the solution then become
\begin{align}
V= 2(\tilde\alpha_1+\tilde \alpha_2) \ t^{-2} \qquad , \qquad
\frac{\partial V}{\partial \Phi_i}=2(\tilde\alpha_1+\tilde \alpha_2)\lambda_{ii} \ t^{-2}
\ . \label{e36}
\end{align}
One arrives at the following system of equations
\begin{align}
&\frac{d(d-1)}{2}a^2=\frac{1}{4}\left(p_1^2+p_2^2\right)+(\tilde \alpha_1+\tilde \alpha_2)
\ , \nonumber \\
&-p_1+d\,b\,p_1=-2(\tilde\alpha_1+\tilde \alpha_2) \lambda_{11} \ , \nonumber \\
&-p_2+d\,b\,p_2=-2(\tilde\alpha_1+\tilde \alpha_2) \lambda_{22} \ , \nonumber \\
&\lambda_{11}p_1+\lambda_{22}p_2 = - 2 \ . \label{e37}
\end{align}
The exact solution is then determined to be
\begin{align}
&p_1=-\frac{2\lambda_{11}}{\lambda_{11}^2+\lambda_{22}^2} \quad , \quad p_2=-\frac{2\lambda_{22}}{\lambda_{11}^2+\lambda_{22}^2} \ , \nonumber \\
&b=\frac{1}{d} \frac{\lambda_c^2}{\lambda_{11}^2+\lambda_{22}^2} \quad , \quad \tilde \alpha_1 +\tilde\alpha_2= \frac{\lambda_c^2-\lambda_{11}^2-\lambda_{22}^2}{(\lambda_{11}^2+\lambda_{22}^2)^2}
\ . \label{e38}
\end{align}
describing two descending scalars. Notice that only one of the integration constants $\Phi_{0k}$ is determined in this case.

%%%%%%%%%%%%%%%%%%%%%%%%%%%%%%%%%%%%%%%%%%%%%%%%%%%%%%%%%%%%%%%%%%%%%%%%%%%%%%%%%%%%%%%%%%%%%%%%%%%%%%%%%%%%%%%%%%%%%%%%%%%%%%%%%
\section{Steep stabilizing potentials and oscillatory behavior near big-bang}

Let us consider the class of potentials
\be
V \ = \ 2 \ \alpha_1 e^{\lambda_1 \Phi} \ + \ 2 \ \alpha_2 e^{- \lambda_2 \Phi} \ , \label{nbb1}
\ee
with $\alpha_i > 0$, that are stabilizing the scalar at a minimum with positive vacuum energy. Whereas the late time dynamics of the scalar towards the minimum is transparent, the early dynamics close to big-bang is interesting. Indeed, according
to the viewpoint going backwards in time, for subcritical exponents $\lambda_i$
the field oscillates until it goes to infinity on one side of the potential. This
can be either of them if both $\lambda_1,\lambda_2 < \lambda_c$ are subcritical or, if
$\lambda_1 > \lambda_c$, $\lambda_2 < \lambda_c$, the field escapes to infinity at the big-bang on the flatter side $\Phi \to - \infty$. If both exponents are supercritical $\lambda_1,\lambda_2 > \lambda_c$, for
any velocity of the scalar approaching big-bang, we expect that the potential will always stop the scalar and send it back into the potential, producing wilder and
wilder oscillations that continue up to the singularity. The goal of this Section
is to study in more detail this intuitive picture in a simple enough toy model.

With the cosmological ansatz (\ref{sod1}) and the asymptotic Kasner solutions of the form (\ref{sod3}), the equations of motion (\ref{sod2}) become
\begin{equation}
\begin{split}
&\frac{d(d-1)}{2}\frac{b^2}{t^2} \ = \ \frac{p^2}{4t^2}+\tilde \alpha_1 \ t^{\lambda_1 p} + \tilde \alpha_2 \ t^{-\lambda_2 p} \ , \\
&\frac{d(d-1)}{2}\frac{b^2}{t^2}-\frac{(d-1)b}{t^2} \ = \ - \frac{p^2}{4t^2} + \tilde \alpha_1 \ t^{\lambda_1 p} +
\tilde \alpha_2 \ t^{-\lambda_2 p} \ , \\
& \frac{(d\,b-1)\, p}{t^2} \ = \ - 2 \ (\lambda_1 \tilde \alpha_1 \ t^{\lambda_1 p} - \lambda_2 \tilde \alpha_2 \ t^{-\lambda_2 p}) \ , \\
\end{split} \ , \label{nbb2}
\end{equation}
where $\tilde \alpha_i = {\alpha}_i e^{\lambda_i \Phi_0}$.
As before, Kasner type solutions are the ones where the terms arising from the scalar potential can be neglected (\ref{sod5})
with respect to the ones arising from the Einstein tensor in the limit $t\rightarrow 0$. In our case this translates to the condition
\begin{equation}
|\lambda_i p| \ < \ 2 \ . \label{nbb3}
\end{equation}
The Kasner coefficients $a$ and $p$ are determined again by (\ref{sod8}) and there is again a critical coefficient $\lambda_c$ defined
in (\ref{sod9}).

There are now three distinct cases of interest:
\begin{itemize}
\item $\lambda_1, \lambda_2 < \lambda_c$.

In this case there are Kasner solutions with the scalar starting on either side ($\pm \infty$) of the minimum. The potential is negligible near big-bang. There are also corresponding particular solutions of the type (\ref{sod10})-(\ref{sod11}), with
$\lambda_1 p = -2$ and $\lambda_2 p = 2$. In this
case however they are not exact solutions anymore, but are valid only near big-bang.
\item $\lambda_1 < \lambda_c$ , $\lambda_2 > \lambda_c$ (or the other way around).

In this case there are Kasner solutions with the scalar starting near big-bang necessarily on the flat side ($- \infty$ for this example) of the potential. This is the typical example of climbing, investigated in a related example in \cite{dkps}, with two exponentials of the same sign of exponent. There is also a LM like
solution (\ref{sod10})-(\ref{sod11}) for $\lambda_1 p = -2$, valid for early time.

\item $\lambda_1, \lambda_2 > \lambda_c$.

There is obviously a de Sitter solution with the scalar sitting at the minimum of
the potential,
\be
A \ = \ \sigma \ t \ , \ \Phi = \Phi_0 \quad , \quad {\rm where} \quad d (d-1) \lambda^2 = V_0 \ , \ V'(\Phi_0)=0
\ , \label{nbb4}
\ee
leading to a de Sitter exponential inflation, free of big-bang singularities.
As expected, in all the three cases above there are solutions evolving exponentially fast to de Sitter one. This can be checked by linearizing around the de Sitter solution
\be
A\ = \ \sigma\ t\ +\ x\ \ , \ \Phi \ = \ \Phi_0 + \phi \ , \label{nbb5}
\ee
with $x,\phi$ suitably small. Notice that since $(d-1) \ddot A = - (1/2) {\dot \Phi}^2$, then $x \sim \phi^2 $. In this case, the leading order eqs. of motion (\ref{sod2}) in the perturbations $x,\phi$ become
\ba
&& d(d-1)\, \sigma\, \dot x \ = \ \frac{1}{4} ({\dot \phi}^2 +  \phi^2 V^{''}_0) \ , \nonumber \\
&& d(d-1)\, \sigma\, \dot x + (d-1) \ddot x \ = \ \frac{1}{4} (- {\dot \phi}^2 +  \phi^2 V^{''}_0) \ , \nonumber \\
&& \ddot \phi + d\, \sigma\, \dot \phi + \phi V^{''}_0 \ = \ 0 \ . \label{nbb6}
\ea
The solutions to (\ref{nbb6}) are
\be
\phi \ = \ \alpha_{\pm} \ e^{\mu_{\pm} t} \ , \ x \ = \ \beta_{\pm} \ e^{2 \mu_{\pm} t} \ , \label{nbb7}
\ee
where
\be
\mu_{\pm} \ = \ \frac{1}{2} (-d\ \sigma \pm \sqrt{d^2\, \sigma^2 - 4 V^{''}_0} ) \ , \ 8 (d-1) \beta_{\pm} \ = \ - \ \alpha_{\pm}^2
\ .  \label{nbb8}
\ee
Notice that $\mu^{\pm} < 0$, so both solutions tend exponentially fast to the de Sitter, as expected. The solutions (\ref{nbb7}),(\ref{nbb8}) are clearly valid only for late enough time, when the perturbations $x,\phi$ are small. For the potential under consideration, $V^{''}_0 / V_0 = \lambda_1 \lambda_2$ and therefore
\be
\Delta \ \equiv \ d^2\, \sigma^2 \ - \ 4 V^{''}_0 \ = \ \left(\frac{d}{d-1} - 4 \lambda_1 \lambda_2\right) V_0 \ . \label{nbb9}
\ee
Therefore,  \\
$\bullet$ for $\lambda_1 \lambda_2 \leq \frac{d}{4(d-1)} = \frac{1}{8} \lambda_c^2$, the scalar is exponentially damped to the minimum. \\
$\bullet$ for $\lambda_1 \lambda_2 > \frac{d}{4(d-1)}$, the scalar is damped, but it also subject to oscillations around the minimum.

\end{itemize}
The third case above $\lambda_1, \lambda_2 > \lambda_c$ is puzzling. There are no possible Kasner solutions near big-bang: the scalar potential with Kasner ansatz is too singular for any initial position of the scalar.
We were also unable to find solutions with  definite limiting behavior
for the scalar field. In Kasner solutions described in previous Sections, starting from a given time and going backwards towards big-bang, the scalar field oscillates with larger and larger amplitudes, until a time close to big-bang where the scalar goes at infinity on the flat side of the potential. In the present example, the potential is too steep to allow this to happen.
Going backwards to big-bang, oscillations become larger and larger and the field continues to oscillates wildly until the singularity is reached in finite time.
This is maybe connected somehow to the singularity theorems \cite{singtheorems}: when the scalar potential becomes more singular than the kinetic terms, the dominant and the strong energy conditions are violated. This oscillatory behavior of the scalar field is reminiscent of the BKL chaotic approach of the singularity in general relativity \cite{bkl} in Kasner solutions.

Whereas we did not find an analytical method to investigate the dynamics, we believe that a useful insight could come from a simplified toy model, with stabilizing scalar potential with critical slope on both sides of the minimum \footnote{This example was worked out in collaboration with  A. Sagnotti. We are grateful to him for numerous discussions which shaped the arguments and conclusions below.}
\begin{equation}
V (\Phi) \ = \ 2 \alpha \ e^{\lambda_c |\Phi|} \ , \label{toy1}
\end{equation}
which is an exponential potential with critical slope on both sides of its minimum, here
at $\Phi=0$. This potential is very similar to his smooth cousin $V (\Phi) \ = \ 4 \alpha \ ch (\lambda_c \Phi)$; we expect the conclusions below to apply to the smooth case of well.
The solution in this case for $\Phi >0$ and $\Phi < 0$ is known analytically in a parametric time $\tau$ and
is given, up to a multiplicative constant, by \cite{dks}\footnote{Strictly speaking one has to do the rescaling $\varphi_n = \sqrt{\frac{d}{2(d-1)}}\Phi_n$ in order to go to the variables used in \cite{dks}, where the eq. (\ref{toy2}) holds.}
\be
\varphi_n(\tau) \ = \ \varphi_{0n} \ - \ (-1)^n \left[ \frac{1}{2}\ \log(\tau+d_n) \ - \ \frac{1}{4} \ (\tau + d_n)^2 \right] \ , \label{toy2}
\ee
where $n=2k$ corresponds to the region I on the left of the minimum $\varphi < 0$, whereas
$n=2k+1$ corresponds to the region II on the right of the minimum $\varphi > 0$. Matching conditions on the field and its derivative have to be imposed at $\varphi=0$ each time the field passes through its minimum. Our goal
is to go backwards in time ($\tau < 0$) starting from large negative values and approaching
the big-bang $\tau=0$. The behavior of the solution closer and closer to the singularity
will hopefully provide insights on the nature of the solution.
For large oscillations ( large $|\varphi_{0n}|$), they are well approximated by a portion of parabola to the right and an almost vertical line to the left, where the argument of log vanishes. So one of the zeroes is for $\tau \approx -d_n$ and the other is far away, and as a result from the zeroes one can find the relations
\be
\varphi_{0n} \ \approx \ \frac{1}{2} \ (-1)^n \ \log(\tau_n+d_n) \ , \qquad \varphi_{0n} \ \approx \ - \ \frac{1}{4} \ (-1)^n \ (\tau_{n-1}+d_n)^2 \ . \label{toy3}
\ee
Moreover, matching the derivatives of $\varphi_n$
\be
\varphi_n^\prime(\tau) \ = \ - \ \frac{1}{2} \ (-1)^n \left[ \frac{1}{(\tau+t_n)} \ - \ (\tau+\tau_n) \right] \label{toy4}
\ee
and $\varphi_{n+1}$ at $\tau_n$ gives the relations
\be
x_n \ \equiv \ \tau_n\ +\ d_n \ = \ \frac{1}{\tau_n\ +\ d_{n+1}} \  . \label{toy5}
\ee
Combining eqs. (\ref{toy3}) now gives
\be
\tau_n+d_n \ \approx \ e^{\,-\, \frac{1}{2}\, (\tau_{n-1}+d_n)^2} \ , \label{toy6}
\ee
so that using eq. (\ref{toy5}) one arrives at the sequence
\be
x_n \ \approx \ e^{\,-\, \frac{1}{2 x_{n-1}^2}} \ , \label{toy7}
\ee
which converges extremely rapidly to zero, and then the peaks of the oscillatory motion are essentially given by
\be
\varphi_{0n} \ \approx \ \frac{1}{2} \ (-1)^n \ \log x_n \ \approx \ - \ \frac{(-1)^n}{4\, x_{n-1}^2} \ . \label{toy8}
\ee

One can also estimate the half--periods in parametric time
\be
\frac{1}{2} \ \Delta \tau_n \ \approx \ - (\tau_n - \tau_{n-1}) \ = \ -(\tau_n+d_n) \ + \ (\tau_{n-1}+d_n) \ = \ - \ x_n \ + \ \frac{1}{x_{n-1}} \ , \label{toy9}
\ee
or
\be
\frac{1}{2} \ \Delta \tau_n \ \approx \  \frac{1}{x_{n-1}} \ . \label{toy10}
\ee
Now turn to cosmic time, using the relation
\be
d t \sim \frac{1}{\sqrt{V}}\ d \tau \ , \label{toy11}
\ee
and let us make the approximation that most of the time within a lob $\varphi \sim \varphi_{0n}$. Then, taking into account the symmetry of the potential $\varphi \to - \varphi$, so that opposite lobs contribute the same, we have
\begin{eqnarray}
&& \frac{1}{2} \ \Delta T_{2k} = \int_{\tau_{2k}}^{\tau_{2k+1}} \frac{d \tau}{ \sqrt{V (\varphi) }}
=  \int_{\tau_{2k}}^{\tau_{2k+1}} \frac{d \tau}{|\tau + d_{2k}|^{1/2} } \ e^{- \frac{1}{4}
(\tau_{2k+1} + d_{2k})^2 +  \frac{1}{4} (\tau + d_{2k})^2 } \nonumber \\
&& =   e^{- \frac{1}{4} (\tau_{2k+1} + d_{2k} )^2 } \int_{x_{2k}}^{\frac{1}{x_{2k+1}}}
\frac{d z}{\sqrt{z}} \ e^{\frac{1}{4} z^2} \simeq \frac{1}{2}
 \ e^{- \frac{1}{4 x_{2k+1}^2}} \int_0^{\frac{1}{x_{2k+1}^2}} dw \ w^{- \frac{3}{4}}
\ e^{\frac{w}{4}} \ . \label{toy12}
\end{eqnarray}
The integral above is well approximated for very small $x_{2k+1}$ by the following function
\begin{align}
 e^{- \frac{1}{4 x_{2k+1}^2}} \int_0^{\frac{1}{x_{2k+1}^2}} dw \ w^{- \frac{3}{4}}
\ e^{\frac{w}{4}} \simeq \left(\frac{1}{x_{2k+1}^2}\right)^{-3/4}
\end{align}
Thus the estimated half period can be written as
\begin{align}
 \frac{1}{2} \ \Delta T_{2k} \simeq \frac{1}{2}x_{2k+1}^{3/2} \rightarrow 0
\end{align}

The upshot is that approaching the singularity, the period of oscillations becomes
smaller and smaller. The oscillations become uncontrollable such that there is no limit
for the scalar field, whereas the scale factor goes monotonically to big-bang. This is to be
contrasted with the Kasner solutions discussed in previous Sections where, after a certain time approaching backwards big-bang, the field would stay in one region (I or II) and go to infinity values at big-bang, as in the example discussed in \cite{dks}.
From this point of view, the de-Sitter solution with the scalar at the minimum, is
unstable (evolving backwards towards the singularity) to small perturbations towards the wild oscillatory behavior described above.

Whereas we do not have concrete examples of controllable setups leading to super-critical potentials of this type, it is easy to
find non-supersymmetric examples in the limit of neglecting backreaction. For example\footnote{We thank A. Sagnotti for discussions
leading to this example.}, type IIB orientifolds with five-form fluxes
lead to a scalar potential for the overall volume breathing mode with $\lambda_1=\lambda_2 = 4\lambda_c/3$.

%%%%%%%%%%%%%%%%%%%%%%%%%%%%%%%%%%%%%%%%%%%%%%%%%%%%%%%%%%%%%%%%%%%%%%%%%%%%%%%%%%%%%%%%%%%%%%%%%%%%%%%%%%%%%%%%%%%%%%%%%%%%%%%%%
\section{Higher derivative corrections}

In this section we examine the effect of adding higher order terms to the action of a scalar field with exponential potential moving in a Kasner background with
$SO(d)$ symmetry. The original action was given by
\begin{equation}
S_0 \ = \ \frac{1}{2}\int d^{d+1}x \sqrt{-g}\left[R-\frac{1}{2}(\nabla\Phi)^2-V(\Phi)\right] \ .
\end{equation}
Assuming that the exponent $\lambda$ of the scalar potential is equal or larger than the critical value $\lambda_c$ then the asymptotic solutions to the equations of motion following from the action above were given by
\begin{equation}
\label{solutions}
A = b \ln t \quad , \quad \phi = p \ln t \qquad {\rm with} \qquad b = \frac{1}{d} \quad , \quad p=\sqrt{\frac{2(d-1)}{d}} \ ,
\end{equation}
describing a climbing scalar and an expanding universe with metric tensor given by
\begin{equation}
\label{metric}
ds^2 \ = \  - dt^2 + e^{2A(t)} \sum_{i=1}^d (dx^i)^2 \ = \ - dt^2 + t^{2/d} \sum_{i=1}^d (dx^i)^2 \ .
\end{equation}
Notice that the Ricci scalar and the kinetic term of $\phi$ behave like $t^{-2}$ when $t\rightarrow0$. We are looking for corrections of order $t^{-(4+qp)}$ which preserve the solutions in eq. (\ref{solutions}).
Such behavior arises from quadratic terms in the curvature that we parametrize in the following way
\begin{equation}
S_1=\frac{1}{2}\int d^{d+1}x \sqrt{-g}e^{-q\Phi}\left[\alpha_1R^2+\alpha_2R_{\mu\nu}R^{\mu\nu}+\alpha_3 R_{\mu\nu\rho\sigma}R^{\mu\nu\rho\sigma}\right] \ .
\label{s1}
\end{equation}
For the particular choice of coefficients $\alpha_1=\eta$, $\alpha_2=-4\eta$, $\alpha_3=\eta$ and $q=0$ the action above is proportional to the Gauss-Bonnet term.
Furthermore, the following terms involving the scalar field $\Phi$
\begin{align}
S_2 &=\frac{1}{2}\int d^{d+1}x\sqrt{-g}e^{-q\Phi}\left[\beta_1 R (\nabla \Phi)^2+\beta_2 R_{\mu\nu}\, \nabla^\mu\Phi\,\nabla^\nu\Phi+\beta_3 (\nabla\Phi)^4+\beta_4(\square\Phi)^2  \right.\nonumber\\
&\left.+ \beta_5 (\nabla \Phi)^2 \square \Phi+\beta_6 R\, \Box\, \Phi+\beta_7 \, R^{\mu\nu}\, \nabla_\mu\nabla_\nu \, \Phi\right] \label{s2}
\end{align}
also have to be included at the $t^{-(4+qp)}$ order. Notice that the degree of singularity of the actions $S_{1,2}$ depends explicitly on the parameter $q$. In string theory, its value is determined by the string perturbation order which can generate these terms in the effective action. For instance, curvature squared terms can arise at disk level \cite{Bachas:1999um} in effective D-brane actions. This corresponds to a value $\tilde q = 1$ in the string frame. In Appendix \ref{string frame}
we show how to translate various quantities from the string frame to the Einstein frame. It follows that we have the following relation
\begin{equation}
q \ = \ \sqrt{\frac{d-1}{8}}\left(\tilde q - 2\frac{d-3}{d-1}\right) \ .
\end{equation}
Then, for the case of a disk contribution we have that $q p \geq-2$ for all $d=1,...,25$ with equality for $d=25$. This implies that the actions $S_{1,2}$ are more singular than $t^{-2}$ for all dimensions smaller than 26.
Similarly, in the case of a one-loop contribution, corresponding to $\tilde q=0$,  we have that $gp\geq -2$ for all $d=1,...,9$ with equality for $d=9$. Hence, $S_{1,2}$ are more singular than $t^{-2}$ for all dimensions smaller than 10.

In general, $\alpha_i,\beta_i$ could be arbitrary functions of $\phi$. In the
string effective models considered in \cite{dks} they are naturally functions of $e^{\Phi}$ such that they become constants when
$e^{\Phi} \rightarrow 0$. The existence of
the climbing phenomenon itself heavily depends on the explicit form of the functions $\alpha_i,\beta_i$, since the descending solutions
correspond to the nonperturbative regime $e^{\Phi} \rightarrow \infty$. This is the critical weak point of the effective field theory analysis: it is very hard to argue for the existence or absence of the descending solutions by including higher-order and higher derivative
corrections to the effective action. What it can be reliably studied however, is the climbing solution and which type of corrections preserve its form. This is therefore the purpose of this section.

One can find in the Appendix the actions $S_1$ and $S_2$ in eqs. (\ref{s1}), (\ref{s2}) written in terms of $A(t)$, $B(t)$ and $\Phi(t)$.
We also show there, that, imposing the absence of higher derivatives and absence of ghosts
in the equations of motion yields the following constraints
\begin{eqnarray}
&&\alpha_1=\eta \qquad \alpha_2=-4\eta \qquad \alpha_3=\eta \label{no-ghost1} \ , \nonumber \\
&&\beta_1+2\beta_2 =0 \qquad \beta_4=\beta_6=\beta_7=0 \ .
\label{no-ghost2}
\end{eqnarray}
Hence, taking into account eqs. (\ref{no-ghost1}) and (\ref{no-ghost2}) we are lead to consider the following action
\begin{eqnarray}
&& S_1+S_2 =  \frac{1}{2}\int d^{d+1}x \sqrt{-g} \ e^{-q\Phi} \Big[\eta \left(R^2-4R_{\mu\nu}R^{\mu\nu}+R_{\mu\nu\rho\sigma}R^{\mu\nu\rho\sigma}\right) \nonumber \\
&& -2\beta_1 G^{\mu\nu}\nabla_\mu \Phi\nabla_\nu\Phi+\beta_3 (\nabla\Phi)^4+\beta_5 (\nabla\Phi)^2\Box\Phi\Big] \ .
\label{sghost}
\end{eqnarray}
Explicitly, in terms of $A(t), B(t)$ and $\Phi(t)$ one has
\begin{equation}
\begin{split}
S_1&= \frac{1}{2}\int d^{d+1}x \ e^{dA-3B-q\Phi} \ d(d-1)(d-2)\eta\left[4 \ddot A \dot A^2 +(d+1)\dot A^4 -4\dot A^3 \dot B\right] \ , \nonumber \\
S_2&=\frac{1}{2}\int d^{d+1}x \ e^{dA-3B-q\Phi}\left[-d(d-1)\beta_1 \dot A^2+\beta_3\dot\Phi^2+\beta_5(\ddot \Phi+d\dot A\dot\Phi-\dot B\dot\Phi)\right]\dot\Phi^2
\ . \end{split}
\end{equation}
Furthermore, we ask that the equations of motion following from the variation of eq. (\ref{sghost}) preserve the (asymptotic) solution in eq. (\ref{solutions}), i.e. the terms
more singular than $t^{-2}$ vanish. One then obtains the constraints
\begin{eqnarray}
\beta_1&=&0  \ , \nonumber \\
\beta_3&=& \frac{(d-2)\left[-3(d-3)+3q^2 dp^2+q(d+3)p\right]}{12d(d-1)} \ \eta
\ , \nonumber \\
\beta_5&=&\frac{(d-2)\left[2(d-3)-3q^2dp^2-q(5d+3)p\right]p}{2d(d-1)(qp+2)} \ \eta \ . \label{q3}
\end{eqnarray}
Notice that we obtain a unique combination of terms up to an overall normalization factor $\eta$, written explicitly in Appendix A.
For the simple case of $q=0$, we find that the combination of terms of order $t^{-4}$ preserving the asymptotic solution in eq. (\ref{solutions}) is
\begin{equation}
\begin{split}
S_1+S_2&=\frac{1}{2}\int d^{d+1}x \ \sqrt{-g} \ \eta \Bigg\{R^2-4R_{\mu\nu}R^{\mu\nu}+R_{\mu\nu\rho\sigma}R^{\mu\nu\rho\sigma}\\
&-\frac{(d-2)(d-3)}{4d(d-1)}\left[(\nabla\Phi)^4-2 \sqrt{\frac{2(d-1)}{d}}(\nabla\Phi)^2\Box\Phi\right]\Bigg\} \ ,
\end{split}
\end{equation}
where $\eta$ is an arbitrary real parameter. Notice that in four dimensions (d=3), for $q=0$, only the Gauss-Bonnet term can be added in order to preserve the solutions in eq. (\ref{solutions}). In the following we will consider a few examples where adding an arbitrary higher order term to the initial Lagrangian will change the behavior of the solution for the scalar field close to the Big Bang, and hence destroying the climbing phenomenon.
%%%%%%%%%%%%%%%%%%%%%%%%%%%%%%%%%%%%%%%%%%%%%%%%%%%%%%%%%%%%%%%%%%%%%%%%%%%%%%%%%%%%%%%%%%%%%%%%%%%%%

%%%%%%%%%%%%%%%%%%%%%%%%%%%%%%%%%%%%%%%%%%%%%%%%%%%%%%%%%%%%%%%%%%%%%%%%%%%%%%%%%%%%%%%%%%%%%%%%%%%%%%%%%%%%%%%%%%
\subsection{Examples of Higher Order Terms which change the behavior of the solution}

In this section we examine the effect of adding arbitrary higher order terms to the original action $S_0$. The first example that we consider is a DBI scalar coupled to gravity. The second one involves adding an extra term of the form
\begin{equation}
G_{\mu\nu} \ \nabla^\mu \Phi \ \nabla^\nu \Phi \ ,
\end{equation}
where the Einstein tensor, $G_{\mu\nu}=R_{\mu\nu}-\frac{1}{2}g_{\mu\nu} R$, is used in order to ensure that the equations of motion do not contain higher derivatives.
\begin{itemize}
\item {\bf Example 1.}
\end{itemize}
We consider a scalar field coupled to gravity with the following DBI-like action
\begin{equation}
S \ = \ \frac{1}{2}\int d^{d+1}x \sqrt{-g} \left[ R-\sqrt{1+(\nabla\Phi)^2}-V(\Phi) \right] \ ,
\end{equation}
that was intensively studied in the recent years as an alternative to slow-roll inflation \cite{dbi}.
The energy-momentum tensor corresponding to the action above is given by
\begin{equation}
T_{\mu\nu} \ = \ \frac{\nabla_\mu\Phi\nabla_\nu\Phi}{2\sqrt{1+(\nabla\Phi)^2}}-\frac{1}{2}g_{\mu\nu}\left[ \sqrt{1+(\nabla\Phi)^2}+V(\Phi)\right] \ .
\end{equation}
Taking into account that we are searching for solutions which are dependent only of time $A=A(t)$, $\Phi=\Phi(t)$, one is lead to the following equations of motion
\begin{equation}
\begin{split}
&\frac{d(d-1)}{2}\dot A^2 \ = \ \frac{1}{2}\left(\sqrt{1-\dot\Phi^2}+\frac{\dot\Phi^2}{\sqrt{1-\dot\Phi^2}}\right)+\frac{1}{2} V(\Phi)
\ , \\
&\frac{d(d-1)}{2}\dot A^2+(d-1)\ddot A \ = \ \frac{1}{2}\sqrt{1-\dot\Phi^2}+\frac{1}{2} V(\Phi) \ , \\
&e^{-dA}\frac{d}{dt}\left(e^{dA}\frac{\dot\Phi}{\sqrt{1-\dot\Phi^2}}\right)+\frac{\partial V}{\partial\Phi} \ = \ 0 \ . \\
\end{split}
\end{equation}
In order to compensate for the l.h.s. (which behaves like $t^{-2}$ for $A= b \ln t$) in the Einstein equations one requests the limiting behavior
\begin{equation}
\frac{\dot\Phi^2}{\sqrt{1-\dot\Phi^2}} \ \simeq \ t^{-2} \qquad {\rm for\ } \quad t \ \rightarrow \ 0 \ .
\end{equation}
The asymptotic solution for the scalar field has then to satisfy
\begin{equation}
\dot\Phi^2 \ = \ 1-p^2t^4 \qquad {\rm which \ implies} \qquad \dot\Phi \ \simeq \ \epsilon \left(1-\frac{p^2t^4}{2}\right) \ ,
\end{equation}
where $\epsilon$ is an undetermined sign. Furthermore, asymptotically we get a linear behavior for $\Phi$:
\begin{equation}
\Phi \ \simeq \ \Phi_0 + \epsilon \left(t-\frac{p^2t^5}{10}\right) \ .
\end{equation}
Finally, plugging everything back in the equations of motion we obtain that the constants $b, p,$ determining the solutions for the scale factor and the scalar field are given by
\begin{equation}
b \ = \ \frac{2}{d} \qquad , \qquad |p| \ = \ \frac{d}{4(d-1)} \ .
\end{equation}
Notice that the scalar potential $V(\Phi) \sim e^{\lambda\Phi}$ is now regular for both solutions found above, hence it does not give any other constraints on the allowable solutions. This is in contrast to the case with standard kinetic term for the scalar field where for $\lambda>\lambda_c$ the descending scalar solution did not exist due to the singularity in the potential term. The conclusion is that a DBI scalar does not present the phenomenon of climbing. The intuitive reason is that the speed of the scalar field is slower due to the
DBI kinetic term and approaches its maximum, light speed value at the big-bang. Because of the slower speed, the potential term is less singular, such that both speed directions are compatible now with field equations.
%%%%%%%%%%%%%%%%%%%%%%%%%%%%%%%%%%%%%%%%%%%%%%%%%%%%%%%%%%%%%%%%%%%%%%%%%%%%%%%%%%%%%%%%%%%%%%%%%
\begin{itemize}
\item {\bf Example 2.}
\end{itemize}
In this example we consider the following action as the starting point \cite{higherder}
\begin{equation}
S = \frac{1}{2} \int d^{d+1}x\sqrt{-g} \left[R-\frac{1}{2}(\nabla\Phi)^2+\frac{1}{2}G^{\mu\nu}\nabla_\mu\Phi\nabla_\nu\Phi - V(\Phi)\right]
\ .
\end{equation}
The Einstein tensor $G_{\mu\nu}$ appearing above is defined in the usual way in terms of the Ricci tensor $R_{\mu\nu}$ and the curvature scalar $R$
\begin{equation}
G_{\mu\nu}\, =\, R_{\mu\nu}-\frac{1}{2}g_{\mu\nu} R \, .
\end{equation}
For a metric background of the form in eq. (\ref{metric}) on obtains the following equations of motion
\begin{equation}
\begin{split}
&\frac{d(d-1)}{2}\dot A^2 \ = \ \frac{1}{4}\dot\Phi^2(1+d^2\dot A^2)+\frac{1}{2}V(\Phi) \ , \\
&e^{-dA}\frac{d}{dt}\left[e^{dA}\dot\Phi(1+d\dot A^2)\right] \ = \ -\frac{\partial V}{\partial\Phi} \ . \\
\end{split}
\end{equation}
Examination of the first equation indicates that the solution for $\Phi$ cannot be singular in order to compensate for the l.h.s. which behaves like $t^{-2}$. One has actually to impose that
\begin{equation}
\dot\Phi^2 \ = \  p^2 \ ,
\end{equation}
thus obtaining that asymptotically
\begin{equation}
\Phi \ = \ \pm \ |p| \ t \ + \ \Phi_0 \ .
\end{equation}
It easy to see that in order to satisfy the equations of motion one must have
\begin{equation}
b \ = \ \frac{2}{d} \qquad , \qquad p \ = \ \pm\sqrt{\frac{2(d-1)}{d}} \ .
\end{equation}
The solution for the scalar field is no longer singular when $t\rightarrow0$, thus the potential term can always be neglected close to the Big Bang. As for the case of a DBI scalar we obtain that adding a term $G^{\mu\nu}\nabla_\mu\Phi\nabla_\nu\Phi$ to the action destroys the phenomenon of climbing, due to the slow-down of the scalar field speed close to the big-bang.

%%%%%%%%%%%%%%%%%%%%%%%%%%%%%%%%%%%%%%%%%%%%%%%%%%%%%%%%%%%%%%%%%%%%%%%%%%%%%%%%%%%%%%%%%%%%%%%%%%%%%%%%%%%%%%
\section{Kasner solutions and climbing : Multiple Scalar Fields}

In this section we generalize the asymptotic analysis of the field equations to the cases with multiple scalar fields propagating in a $SO(d)$ symmetric gravitational background with the potential already considered previously
\begin{equation}
V(\Phi_1, ..., \Phi_n) \ = \ 2 \sum_i \alpha_i \ e^{\sum_j\lambda_{ij}\Phi_j} \ ,
\label{potential}
\end{equation}
where we have introduced the positive constants $\alpha_i, \lambda_{ij}>0$. The assumption that all $\lambda_{ij}$'s are positive ensures that the potential above is a monotonically increasing function of $\Phi_k$ for any $k$. As discussed in detail in Section 3, the multi-exponential
potentials allow also for exact solutions, which have the interpretation of generalized Lucchin-Mattarese attractors. The goal is to find conditions for
$\lambda_{ij}$ in order to force one scalar to climb its potential. As we will
see, it is straightforward to find a sufficient condition. The most
general necessary and sufficient condition to be satisfied is more involved and will
be explicitly worked out for the case of two scalars only.
%%%%%%%%%%%%%%%%%%%%%%%%%%%%%%%%%%%%%%%%%%%%%%%%%%%%%%%%%%%%%%%%%%%%%%%%%%%%%%%%%%%%%%%%%%%%%%%%%%%%
\subsection{Kasner solutions with $SO(d)$ Symmetry}

The equations of motion describing $n$ scalar fields $\Phi_1, ..., \Phi_n$ minimally coupled to gravity in the background in eq . (\ref{sod1}) are:
\begin{equation}
\begin{split}
&\frac{d(d-1)}{2} \dot A^2 \ = \ \frac{1}{4}\sum_i \dot \Phi_i^2+\frac{1}{2}V(\Phi) \ , \\
&\frac{d(d-1)}{2}\dot A^2+(d-1)\ddot A \ = \ - \frac{1}{4} \sum_i \dot \Phi_i^2+\frac{1}{2}V(\Phi) \ , \\
&\ddot \Phi_i+d\dot A\dot\Phi_i \ = \ - \frac{\partial V(\Phi)}{\partial\Phi_i} \ . \\
\end{split}
\end{equation}
Neglecting the terms involving the scalar potential in the limit $t \rightarrow 0$ implies that the solutions to the system of equations above have the following asymptotic behavior
\begin{equation}
A \ = \frac{1}{d}  \ln t + A_0\qquad , \qquad \Phi_i \ = \ p_i\ln t +\Phi_0\ .
\label{kassol}
\end{equation}
As before, we have chosen the big-bang time to be $t_0=0$. Moreover, the constants $p_i$ have to satisfy the following Kasner type constraint
 \begin{equation}
\sum_i p_i^2 \ = \ \frac{2(d-1)}{d} \ ,
\label{sphere}
\end{equation}
in order for eq. (\ref{kassol}) to be an asymptotic solution. We now have that the space of (asymptotic) solutions for $n$ scalar fields is parametrized by a $(n-1)$ dimensional sphere. Thus, there is a continuously varying set of climbing and descending solutions. These solutions are valid as long as the constraint in eq. (\ref{sod5}) is satisfied. For the scalar potential in eq. (\ref{potential}) one obtains that we must have the following inequalities
\begin{equation}
\sum_j \lambda_{ij} \ p_j \ > \ - 2 \ .
\end{equation}
 The question we are asking is the following : what is the condition on $\lambda_{ij}$ such that we have at least one scalar climbing?
This is equivalent with asking that the solution in eq. (\ref{sphere}) does not exist in the region where all $p_i$'s are negative.

We partition the space of solutions into $2^n$ regions classified by the possible signs of each $p_i$.
\begin{center}
\begin{tabular}{|c|c|c|c|c|}
  \hline
  \ & $p_1$ & $p_2$ & $...$ & $p_n$ \\ \hline\hline
  I & + & + & $...$ & + \\
  II & + & + & $...$ & - \\
  $...$ & $...$ & $...$ & $...$ & $...$ \\
  $2^n$ & - & - & - & - \\
  \hline
\end{tabular}
\end{center}

For convenience, we rescale the $p_i$'s such that we can work on a sphere or radius $1$. Thus, we make the substitution
\begin{equation}
p_i \ \rightarrow \ \sqrt{\frac{d}{2(d-1)}} \ p_i \ .
\label{rescaling}
\end{equation}
 A given solution $(p_1,...,p_n)\in S^{n-1}\equiv\{(x_1,...,x_n)\in \mathbb{R}^n;\ x_1^2+...+x_n^2=1\}$ exist as long as the following condition is satisfied
\begin{equation}
\sum_j \lambda_{ij}p_j \ > \ - \ \sqrt{\frac{2d}{d-1}} \ .
\end{equation}
where above we have taken into account the rescaling in eq. (\ref{rescaling}).

Let us denote by $\Omega\subset S^{n-1}$ the subset where all $p_i$'s are negative. Then the most general condition in order to have climbing is the following:

There exist indices $i_1,...,i_k \in\{1,...,n\}$ and the subsets $\Omega_{i_1},...,\Omega_{i_k} \subset S^{n-1}$ such that the following two conditions are satisfied:
\begin{equation}
\Omega \ \subset \ \Omega_{i_1} \cup... \cup \Omega_{i_k} \ . \label{condition1}
\end{equation}
and
\begin{equation}
F_{i_m}(p_1,...,p_n) \ \equiv \ \sum_j\lambda_{i_mj} \ p_j \ \leq \ - \sqrt{\frac{2d}{d-1}} \qquad {\rm for\ all} \qquad (p_1,...,p_n) \in \Omega_{i_m} \ , \label{condition2}
\end{equation}
with $m=1,...,k$. Eq. (\ref{condition2}) implies that the solutions corresponding to $(p_1, ...,p_n )$ in the region $\Omega_{i_m}$ violate the constraint in eq. (\ref{sod5}), and thus do not exist as asymptotic solutions. Further, eq. (\ref{condition1}),(\ref{condition2}) entails that all the subset of solutions in $\Omega$, for which all scalars are descending, is not allowed by the scalar
potential considered.

In view of the above, it is easy to see that for the case of $n$ scalar fields, the following condition is sufficient (though not necessary) to have climbing:

There exist an index $i\in\{1,...,n\}$ such that
\begin{equation}
F_i(p_1,...,p_n) \ \leq \ - \sqrt{\frac{2d}{d-1}} \qquad {\rm for\ all} \qquad (p_1,...,p_n) \in \Omega \ .
\label{sufficient}
\end{equation}
 One then can show (by induction) that the condition in eq. (\ref{sufficient}) is equivalent to demanding that
\begin{equation}
\lambda_{i1}, ..., \lambda_{in}\, \geq\, \lambda_c \qquad \textrm{with\ } i\in\{1,...,n\} \textrm{\ a fixed index.}
\end{equation}

%%%%%%%%%%%%%%%%%%%%%%%%%%%%%%%%%%%%%%%%%%%%%%%%%%%%%%%%%%%%%%%%%%%%%%%%%%%%%%%%%%%%%%%%%%%%%%%%%%%%%%%%%%%%%%%%%%%%
\subsubsection{Kasner solutions with two scalar fields}

In the following we will examine explicitly the most general condition for climbing in the case of two scalar fields with monotonically increasing potential in both directions.
In this case, we can parametrize the solutions to $p_1^2+p_2^2=1$ by the angle $\theta\in[0,2\pi)$ in the following way
\begin{equation}
p_1 \ = \ \cos{\theta} \qquad , \qquad p_2 \ = \ \sin{\theta} \ .
\end{equation}
We have four regions depending on the four possible signs of $p_1, p_2$
\begin{center}
\begin{tabular}{|c|c|c|c|}
  \hline
  \ & $p_1=\cos{\theta}$ & $p_2=\sin{\theta}$ & $\theta$  \\ \hline\hline
  I & + & + & $(0,\frac{\pi}{2})$ \\
  II & - & + & $(\frac{\pi}{2},\pi)$ \\
  III & - & - & $(\pi,\frac{3\pi}{2})$ \\
  IV & + & - & $(\frac{3\pi}{2},2\pi)$ \\
  \hline
\end{tabular}
\end{center}
Let us start by analyzing the (stronger) condition
\begin{equation}
F(p_1,p_2)=\lambda_{11} p_1 + \lambda_{12} p_2 = \lambda_{11} \cos{\theta}+
\lambda_{12} \sin{\theta} \ \leq \ - \sqrt{\frac{2d}{d-1}} \qquad {\rm for}\ \theta\in\left(\pi,\frac{3\pi}{2}\right)\equiv {\rm III}
\label{strong}
\end{equation}
In the region III, corresponding to a solution with two descending scalars, the function F has only one extremum point which is a minimum. This implies that the function F, restricted on the open interval $(\pi,\frac{3\pi}{2})$, has a maximum on the boundary, that is, in one of the points $\pi$ and $\frac{3\pi}{2}$:
\begin{equation}
\sup_{\theta\in\left(\pi,\frac{3\pi}{2}\right)}F(\theta) \ = \ \max\left(F(\pi), F\left(\frac{3\pi}{2}\right)\right)= \ \max \ (-\lambda_{11}, -\lambda_{12}) \ .
\end{equation}
\begin{figure}[hc!]
  \begin{center}
  \includegraphics[width=9cm]{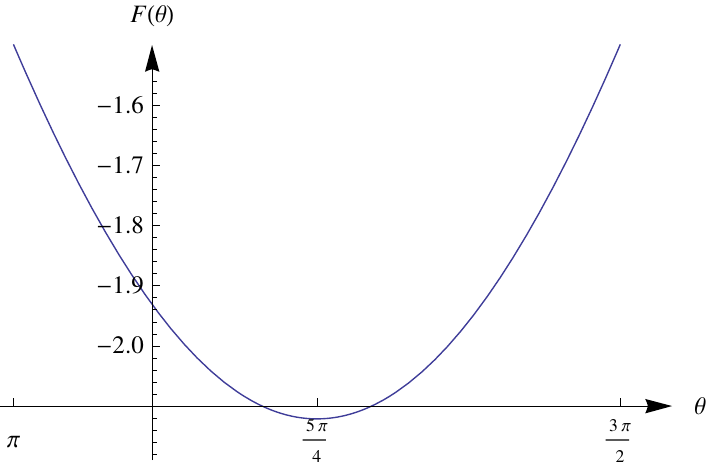}\\
  \caption{We represent the function $F(\theta)=\frac{3}{2}\cos{\theta}+\frac{3}{2}\sin{\theta}$ in the interval $\theta\in(\pi,\frac{3\pi}{2})$. The function has a minimum. Its maximum value is $3/2$ obtained on the ends of the interval.}\label{fig}
  \end{center}
\end{figure}\\
Hence, we have shown that the condition in eq. (\ref{strong}) is equivalent to the following
\begin{equation}
\lambda_{11}, \lambda_{12} \ \geq \ \sqrt{\frac{2d}{d-1}} \ \equiv \ \lambda_c \ .
\label{eqstrong}
\end{equation}
Notice that eq. (\ref{eqstrong}) is not the most general condition to have climbing. Recall that, for two scalars the scalar potential that we consider, evaluated on the
Kasner solution,  is of the
following form
\begin{equation}
V(\Phi_1(t), \Phi_2(t))= \alpha_1 \ t^{F_1(p_1,p_2)}+\alpha_2 \ t^{F_2(p_1,p_2)} \ ,
\end{equation}
with the two functions $F_{1,2}$ being linear in $p_1, p_2$:
\begin{equation}
\begin{split}
&F_1(p_1,p_2) \ = \ \lambda_{11}p_1 \ + \ \lambda_{12}p_2 \ , \\
&F_2(p_1,p_2) \ = \ \lambda_{21}p_1 \ + \ \lambda_{22}p_2 \ . \\
\end{split}
\end{equation}
In general, we can have climbing if $F_1<-\sqrt{\frac{2d}{d-1}}$ in the interval $\Omega_1\subset (\pi,\frac{3\pi}{2})$ and that $F_2<-\sqrt{\frac{2d}{d-1}}$ in the interval $\Omega_2\subset(\pi,\frac{3\pi}{2})$ where $\Omega_1\cup \Omega_2 = (\pi,\frac{3\pi}{2})$.

It is not difficult to show that, in the general case, one has a climbing scalar if and only if one of the following conditions is satisfied:
\begin{enumerate}[\qquad (a)]
\item $\lambda_{11} \ , \ \lambda_{12} \ \geq \ \lambda_c$ \qquad ,
\item $\lambda_{21} \ , \ \lambda_{22} \ \geq \ \lambda_c$ \qquad ,
\item $\lambda_{11} \ , \ \lambda_{22} \ \geq \ \lambda_c$ \ and \ $\theta_1^* \ \geq \ \theta_2^*$ \qquad ,
\item $\lambda_{12} \ , \ \lambda_{21} \ \geq \ \lambda_c$ and $\tilde\theta_2^* \ \geq \ \tilde\theta_1^*$ \qquad ,
\end{enumerate}
where the angles $\theta_{1,2}^*$ are given by
\begin{align}
&\tan \theta_1^* \, =\,  \frac{1}{\lambda_{12}^2-\lambda_c^2} \left(
- \lambda_{11} \lambda_{12} - \lambda_c \sqrt{\lambda_{11}^2+ \lambda_{12}^2-\lambda_c^2}\, \right)
\ , \\
& \tan \theta_2^* \, =\, \frac{1}{\lambda_{22}^2-\lambda_c^2} \left(
- \lambda_{21} \lambda_{22} + \lambda_c \sqrt{\lambda_{21}^2+ \lambda_{22}^2-\lambda_c^2}\, \right)
\ .
\end{align}

Similar expressions can be written down for $\tilde\theta_{1,2}^*$ by flipping the signs inside the parentheses.
Let us explicitate a bit more the case (c). One implicitly assumes that we have $\lambda_{12}, \lambda_{21}<\lambda_c$ so that we do not fall on the cases (a) or (b). Then the function $F_1(\theta)\leq -\lambda_c$ when $\theta\in \Omega_1=[\pi,\theta_1^*]$ and $F_2(\theta)\leq -
\lambda_c$ when $\theta\in \Omega_2=[\theta_2^*,\frac{3\pi}{2}]$. Thus the condition $\Omega_1\cup \Omega_2=\left[\pi,\frac{3\pi}{2}\right]$ corresponds to having $\theta_1^*\geq\theta_2^*$. Notice that the angles $\theta_1^*, \theta_2^*$ are solutions to the equations:
\begin{equation}
F_1(\theta_1^*) \ = \ F_2(\theta_2^*) \ = \ \lambda_c \ .
\end{equation}
They are unique (in the third quadrant) provided that we have $\lambda_{11}, \lambda_{22} \, \geq\,  \lambda_c$ and $\lambda_{12}, \lambda_{21}\, <\, \lambda_c$.
%%%%%%%%%%%%%%%%%%%%%%%%%%%%%%%%%%%%%%%%%%%%%%%%%%%%%%%%%%%%%%%%%%%%%%%%%%%%%%%%%
\subsubsection{Two Scalar Fields with Potential Monotonic in one Direction}
\label{stabilizant}

Up to now we have analyzed only scalar potentials that were increasing functions of the corresponding scalar fields. Here we relax this assumption for the case of two scalar fields. Concretely, we consider a model with the following scalar potential
\begin{equation}
V(\Phi_1, \Phi_2) = 2 \alpha_1 e^{\lambda_{11}\Phi_1 +  \lambda_{12} \Phi_2} +
2 \alpha_2 e^{\lambda_{21}\Phi_1-\lambda_{22}\Phi_2} \ ,
\end{equation}
with all $\lambda_{ij} > 0$ by convention, as before.
Thus $V$ is an increasing function with respect to $\Phi_1$ and as a function of $\Phi_2$ it has a minimum. The asymptotic solutions are of the form
\begin{equation}
\Phi_1 = p_1 \ln t + \Phi_{01} \qquad \qquad \Phi_2 = p_2 \ln t +\Phi_{02}\ ,
\end{equation}
with $p_1, p_2$ varying over the one dimensional sphere
\begin{equation}
p_1^2 + p_2^2\, =\, \frac{2(d-1)}{d} \ .
\end{equation}
We can find a condition to have the field $\Phi_1$ climbing irrespective of what the field $\Phi_2$ does. Indeed, this behavior can be obtained if the following conditions hold
\begin{equation}
\begin{split}
&\lambda_{11}p_1 +\lambda_{12} p_2 \, \leq\,  -2 \qquad \textrm{for\ all} \qquad(p_1,p_2)  \in \textrm{III} \ , \\
&\lambda_{21}p_1 -\lambda_{22} p_2 \, \leq\,  -2 \qquad \textrm{for\ all} \qquad (p_1,p_2)\ \in \textrm{II} \ .
\end{split}
\end{equation}
As in the previous section, it is easy to see that the two conditions above are equivalent to have
\begin{equation}
\lambda_{11},\ \lambda_{12}, \ \lambda_{21}, \ \lambda_{22} \ \geq \ \lambda_c \ . \label{suff}
\end{equation}
The condition $\lambda_{11}, \lambda_{12} \geq \lambda_c$ forbids any solution with $p_1<0, p_2<0$ whereas the condition $\lambda_{21},\, \lambda_{22} \geq \lambda_c$ forbids any solution with $p_1<0,\, p_2>0$. Notice that, similar to the case analyzed in Section 6.1.1 , conditions more general than eq. (\ref{suff}) are possible also here.

%%%%%%%%%%%%%%%%%%%%%%%%%%%%%%%%%%%%%%%%%%%%%%%%%%%%%%%%%%%%%%%%%%%%%%%%%%%%%%%%%%%%%%%%%%%%%%%%%%%%%%%%%%%%%
\section{Climbing in effective string models with moduli stabilization}

A particularly interesting framework for discussing the early time dynamics are models with moduli stabilization \cite{gkp,kklt}, with moduli fields, called $M$ describing the fluctuations of the dilaton and internal space in string theory. Since the potentials are of stabilizing type and the kinetic terms are non-canonical, the analysis of climbing is more involved than in the examples in the previous sections.
One should emphasize the lagrangians used below cannot be exact, in particular for small field valued the effective field theory approximation breaks down and a full-fledged string theory approach would be necessary. The conclusions in this
section are based on the (big) assumption that the effective lagrangian describes
well the dynamics for all values of moduli fields. Our goal here is
to write down conditions on the coefficients $n_i$ of the so-called uplift potentials
below (\ref{uplift}) in order for the climbing phenomenon to take place.

Let us consider an effective action containing $N$ scalar fields coupled to gravity
\begin{equation}
{\cal S} \ = \ \frac{1}{2}\int d^{d+1}x\sqrt{-g} \left[R-2K_{i\bar \jmath}(\partial M^i)(\partial \bar M^{\bar \jmath})-V_F(M^k, \bar M^{\bar k}) - V_{up}(Re(M^k)) \right] \ ,
\end{equation}
where $g$ is the metric tensor of $d+1$ dimensional space-time, $\{M^i\}_{i=1,...,N}$ are $N$ complex scalar fields with K\"{a}hler potential given by
\begin{equation}
K \ = \ - \sum_{i=1}^N r_i \ \ln (M^i+\bar M^i) \ ,
\label{kahler}
\end{equation}
with unspecified, for the time being, superpotential $W (M^i)$. Whereas $V_F$ is a scalar potential containing effects
from fluxes and nonperturbative effects, the uplift potential $V_{up}$ can come from
antibranes \cite{kklt}, D-terms generated by magnetic fluxes \cite{duplift} or F-terms from a dynamical supersymmetry breaking sector \cite{fuplift}. The uplift is
not depending on the imaginary parts of $M^i$( axions) and is generically of the following form
\begin{equation}
V_{up} \ = \ \frac{D}{\prod_{i=1}^N (M^i+\bar M^{\bar\imath})^{n_i}} \ , \label{uplift}
\end{equation}
where the integers $n_i$ have different values for the three uplifts mentioned above.
We search for asymptotic (Kasner) solutions to the equations of motion which depend only on time. The metric tensor is assumed to have $SO(d)$ symmetry
\begin{equation}
d s^2 \ = \ -d t^2 \ + \ e^{2A(t)} \ \sum_{i=1}^d (dx^i)^2 \ .
\label{kasner2}
\end{equation}
The Einstein equations for this metric are
\begin{equation}
\begin{split}
&\frac{d(d-1)}{2}\dot A^2 \ = \ K_{i\bar\jmath}\, \dot M^i\dot{\bar M}^{\bar\jmath}+\frac{1}{2}V(M^k,\bar M^k) \ , \\
&\frac{d(d-1)}{2}\dot A^2+(d-1)\ddot A \ = \ - K_{i\bar\jmath}\, \dot M^i\dot{\bar M}^{\bar\jmath}+\frac{1}{2}V(M^k,\bar M^k) \ . \\
\label{einstein}
\end{split}
\end{equation}
In addition, one obtains the following equations of motion for the complex scalar fields $M^k$
\begin{equation}
\ddot M^k + d \dot A \dot M^k+\Gamma^{k}_{ij}\, \dot M^i \dot M^j \ = \ - \frac{1}{2}K^{k\bar l}
\frac{\partial V}{\partial \bar M^{\bar l}} \ ,
\label{scalar}
\end{equation}
where the K\"ahler connection $\Gamma^k_{ij}$ is expressed in terms of the K\"ahler potential as follows
\begin{equation}
\Gamma^{k}_{ij} \ = \ K^{k\bar l} K_{ij\, \bar l} \ .
\end{equation}
 Adding and subtracting the two equations in (\ref{einstein}) one obtains
\begin{equation}
\begin{split}
& d(d-1)\dot A^2 + (d-1)\ddot A \ = \ V \ , \\
&-(d-1)\ddot A \ = \ 2 \ K_{i\bar\jmath}\, \dot M^i\dot{\bar M}^{\bar\jmath} \ . \\
\label{einstein2}
\end{split}
\end{equation}
We first neglect in the equations of motion the terms containing the scalar potential. The solutions that we find have to satisfy the analogous conditions in eq.(\ref{sod5}). In this case,
integration of the first equation leads to the following solution (up to integration constants) for the scale factor :
\begin{equation}
A \ = \ b \ln t \qquad {\rm with}\qquad  b \ = \ \frac{1}{d} \ .
\end{equation}
It is useful to parameterize the complex scalar fields $M^k$ in the following way
\begin{equation}
M^k \ = \ e^{\frac{\Phi_k}{\sqrt{r_k}}} \ + \ i \frac{\theta_k}{\sqrt{r_k}} \ .
\end{equation}
Then making use of the K\"ahler potential in eq. (\ref{kahler}) one obtains that the kinetic terms for the fields $M_k$ can be written as
\begin{equation}
K_{i\bar\jmath}\, \dot M^i\dot{\bar M}^{\bar\jmath} \ = \ \frac{1}{4}\sum_{i=1}^N \left( \dot \Phi_k^2 +
e^{-\frac{2\Phi_k}{\sqrt{r_k}}}\, \dot\theta_k^2 \right) \ .
\end{equation}
We search for solutions of the following form
\begin{equation}
\Phi_k(t) \ = \ p_k\ln(t) \qquad , \qquad \dot\theta_k(t) \ = \ q_k \ t^{\frac{p_k}{\sqrt{r_k}}-1} \ .
\label{kscalar}
\end{equation}
The exponent for the axion $\theta_k$ has been chosen in such a way that the corresponding kinetic term behaves like $t^{-2}$. This is due to the fact that the parametrization for the real and imaginary part of $M_k$ is different. Plugging everything in eq. (\ref{einstein2}) one obtains the Kasner sphere condition
\begin{equation}
\sum_{k=1}^N \ \left( p_k^2 + q_k^2 \right) \ = \ \frac{2(d-1)}{d} \ .
\end{equation}
Rewriting the equations of motion for $M^k$ in terms of the real fields $\Phi_k, \theta_k$ and making use of the K\"ahler potential in eq. (\ref{kahler}) one is led
to the field equations
\begin{equation}
\begin{split}
&\ddot\Phi_k+d\dot A\dot\Phi_k+\frac{1}{\sqrt{r_k}}e^{-\frac{2\Phi_k}{\sqrt{r_k}}}\dot\theta_k^2 \ = \ - \frac{\partial V}{\partial \Phi_k} \ , \\
&e^{-\frac{\Phi_k}{\sqrt{r_k}}}\left(\ddot \theta_k+d\dot A\dot\theta_k -\frac{2}{\sqrt{r_k}}\dot\theta_k\dot\Phi_k\right) \ = \ -e^{\frac{\Phi_k}{\sqrt{r_k}}}\frac{\partial V}{\partial \theta_k} \ . \\
\end{split}
\end{equation}
Neglecting the terms involving the scalar potential and plugging the solutions in eq. (\ref{kscalar}), one gets
\begin{equation}
p_k (-1+d\,b) + \frac{q_k^2}{\sqrt{r_k}} \ = \ 0 \quad , \quad q_k\left(-\frac{p_k}{\sqrt{r_k}}-1 + da \right) \ = \ 0 \ .
\end{equation}
Further making use of the fact that $b=1/d$ it follows immediately we must have
\begin{equation}
q_k \ = \ 0 \qquad {\rm for \ all \ } \qquad k=1,...,N \quad \ ,
\end{equation}
i.e. the axions dynamics is frozen close to the big-bang.
To summarize, the asymptotic solutions for the scalar fields $\Phi_k=p_k\ln(t)$, $\dot\theta_k=q_k t^{\frac{p_k}{\sqrt{r_k}}-1}$ are parameterized by the Kasner sphere
\begin{equation}
\sum_{k=1}^Np_k^2\, =\, \frac{2(d-1)}{d} \qquad , \qquad q_k \, = \, 0 \ ,
\end{equation}
subject to the constraints arriving from the behavior of the scalar potential near the singularity $t=0$
\begin{equation}
V \quad , \quad  \frac{\partial V}{\partial \Phi_k} \quad , \quad e^\frac{\Phi_k}{\sqrt{r_k}}\frac{\partial V}{\partial \theta_k}
\ \sim \ O(t^{-2+\epsilon}) \qquad {\rm with\ } \ \epsilon > 0 \ .
\label{regularity}
\end{equation}
In order to analyze the conditions above one needs to specify an explicit form of the scalar potential.
%%%%%%%%%%%%%%%%%%%%%%%%%%%%%%%%%%%%%%%%%%%%%%%%%%%%%%%%%%%%%%%%%%%%%%%%%%%%%%%%%%%%%%%%%%%%%%%%%%%%%%%%%%%%%%%%%%
\subsection{Climbing in KKLT}
\label{seckklt}

We consider the KKLT scenario for moduli stabilization  \cite{kklt} in type IIB strings. It is a particular case of the general setting considered above with one complex scalar field in $d=3$ space dimensions
\begin{equation}
T \, =\,  e^{\frac{\Phi}{\sqrt{3}}} + i \frac{\theta}{\sqrt{3}} \ ,
\end{equation}
with the following K\"ahler and superpotential functions
\begin{equation}
K \, =\, - 3 \ \ln (T+\bar T) \ , \qquad \ W\, =\,  W_0 + c \ e^{- {\tilde b} T} \ . \label{kklt1}
\end{equation}
The asymptotic solution that one finds when neglecting the scalar potential terms in the equations of motion close to the singularity $t\rightarrow 0$ are
\begin{equation}
A\, =\,  \frac{1}{3} \ln t \quad , \quad \Phi\, =\, \pm \frac{2}{\sqrt{3}} \ln t \quad , \quad
\theta = 0 \ . \label{kklt2}
\end{equation}
In order to examine the existence of the descending solution we consider the following scalar potential uplift depending only on $\Phi$
\begin{equation}
V_{up}\, =\, \frac{D}{(T+\bar T)^{n_T}}\, =\, \frac{D}{2^{n_T}}e^{- \frac{n_T}{\sqrt{3}}\Phi} \ . \label{kklt3}
\end{equation}
Notice that in the KKLT scenario there is also a supergravity F-term ($V_F$) part of the scalar potential containing the dependence on the axion field $\theta$. This term is exponentially suppressed for large $Re \ T$, whereas for small $Re \ T$ it behaves as $(T + {\bar T})^{-2}$, although it should not really be trusted.  As it can be checked aposteriori, even in the small field limit $Re \ T \to 0$, one can neglect this part
close to big-bang $t\rightarrow 0$, since the condition for climbing will impose the uplift to dominates the potential both for large and small field values. We are then reduced to the situation of one real scalar field with exponential potential $V\sim e^{- \lambda \Phi}$ defined by the constant
\begin{equation}
\lambda \ = \ \frac{n_T}{\sqrt{3}} . \label{kklt4}
\end{equation}
Using the fact that the critical coefficient $\lambda_c$ in $d=3$ is given by
\begin{equation}
\lambda_c \ = \ \sqrt{3} \ , \label{kklt5}
\end{equation}
one obtains that the condition to have climbing in the KKLT case is
\begin{equation}
n_T \ \geq \ 3 \ . \label{kklt6}
\end{equation}
Notice that the antibranes uplift \cite{kklt} generates a subcritical slope $n_T=2$, whereas
the D-terms \cite{duplift} and F-terms \cite{fuplift} uplifts generate precisely the critical value
$n_T=3$.

%%%%%%%%%%%%%%%%%%%%%%%%%%%%%%%%%%%%%%%%%%%%%%%%%%
\subsection{Quasi-exact Solutions in KKLT}

We consider solutions in the KKLT scenario where the uplift potential behaves like $V_{up}\sim t^{-2}$, by still ignoring the non-perturbative F-term potential. As such, unlike solutions described in Section 3, these solutions are not exact, but approximate on the portion of the potential where nonperturbative effects are negligible.  We also assume the same behavior for the various terms in the equations of motion. This implies that the solution of the scale factor $A$, the scalar field $\Phi$ and the axion field $\theta$ is of
the following form
\begin{align}
A(t)\ = \ b \ln t \qquad ,\qquad \Phi(t)\ =\ \frac{2\sqrt{3}}{n_T}\ln t +\Phi_0 \qquad , \qquad \dot\theta(t)\ =\ q\ t^{\frac{2-n_T}{n_T}}\, .
\label{sol KKLT}
\end{align}
Notice that the system considered here is different from the one analyzed in Section 3 as now we have that the scalar potential does not depend on the axion field $\theta$.
\begin{equation}
V_{up}(\Phi)\ =\ 2\, \alpha\, e^{-\lambda \Phi} \qquad , \qquad \alpha \ = \ \frac{D}{2^{n_T+1}} \ .
\label{pot KKLT}
\end{equation}
Replacing eqs. (\ref{sol KKLT}), (\ref{pot KKLT}) into the equations of motion for the scale factor $A(t)$, scalar field $\Phi(t)$ and axion $\theta(t)$ given by
\begin{align}
&3 \dot A^2\ =\ \frac{1}{4}\left(\dot\Phi^2+e^{-\frac{2\Phi}{\sqrt{3}}}\dot\theta^2\right)+\frac{1}{2}V \ , \nonumber \\
&3 \dot A^2 +2\ddot A\ =\ -\frac{1}{4}\left(\dot\Phi^2+e^{-\frac{2\Phi}{\sqrt{3}}}\dot\theta^2\right)+\frac{1}{2}V \ , \nonumber \\
&\ddot \Phi + d\, \dot A\, \dot\Phi+\frac{1}{\sqrt{3}} e^{-\frac{2\Phi}{\sqrt{3}}}\dot\theta^2\ =\ -\frac{\partial V}{\partial \Phi} \ , \nonumber \\
&e^{-\frac{\Phi}{\sqrt{3}}}\left(\ddot\theta+d\,\dot A\, \dot\theta-\frac{2}{\sqrt{3}}\,\dot\theta\,\dot\Phi\right)\ =\ e^{\frac{\Phi}{\sqrt{3}}}\frac{\partial V}{\partial \theta}\, ,
\end{align}
one obtains that the constants $b, q$ and $\Phi_0$ are determined as follows (for the case $q\neq 0$)
\begin{align}
b\ =\ \frac{n_T+2}{d\,n_T} \qquad \qquad \tilde \alpha\ =\ \frac{2\,(n_{T}+2)}{3\, n_T^2} \qquad \qquad \tilde q\  =\ \pm \frac{2}{n_T\sqrt{3}}\sqrt{ n_T(n_T+2)-9} \ .
\end{align}
In the above, we have defined the constants $\tilde \alpha$ and $\tilde q$ to include the ``initial" condition $\Phi_0$ in the following way
\begin{align}
\tilde \alpha\ =\ \alpha\, e^{-\frac{n_T}{\sqrt{3}}\Phi_0} \qquad , \qquad \tilde q\ =\ q\, e^{-\frac{\Phi_0}{\sqrt{3}}} \ .
\end{align}
Notice that even though we always have $\tilde \alpha >0$ (since $n>0$), there is a non-trivial positivity condition coming from the expression for $\tilde q$. Namely, one needs to have the following condition satisfied in order for the solutions above to be valid
\begin{equation}
n_T > \sqrt{10}-1 \ \simeq 2.16 \ \ .
\end{equation}
This condition and the resulting solution was first found in \cite{dks} (eqs. (3.12)-(3.13) there), where was interpreted as a late time attractor. Here we remark that the solution is actually exact as far as the nonperturbative piece of the potential is ignored.
In conclusion, there exists a descending solution for $\Phi$ with fixed $\Phi_0$ for any $n_T\geq 3$. Together with the descending solution with free asymptotics found in the previous section which exists for $n_T<3$, it follows that in the KKLT scenario  there is always a descending solution available for the scalar field $\Phi$ irrespective of the value of $n_T$. This exact solution is an attractor for late time for more general solutions. For early time, however,
it exists for a fine-tuned value of the initial condition $\Phi_0$, whereas for the other
initial conditions climbing persists for $n_T \ge 3$. Moreover, the solution is valid only in the region where the nonperturbative portion of the potential is negligible.  \\
Finally, a different exact solution corresponding to Lucchin-Matarrese can be found if one freezes the axion dynamics (i.e. $q=0$).
\begin{align}
A\ = \ \frac{\sqrt{3}}{n_T} \ln t \qquad , \qquad \Phi \ = \ \frac{2\sqrt{3}}{n_T} \ln t + \frac{\sqrt{3}}{n_T}\ln \frac{3(9-n_T^2)}{\alpha n_T^4} \qquad , \qquad \theta\ =\ \theta_0 \ .
\end{align}
This solution describes a descending scalar and it exists if and only if $n_T<3$.

%%%%%%%%%%%%%%%%%%%%%%%%%%%%%%%%%%%%%%%%%%%%%%%%%%%%%%%%%%%%%%%%%%%%%%%%%%%%%%%%%%%%%%%%%%%%%%%%%%%%%%%%%%%%%%%%%%%%%%%%%%%%%%%
\subsection{Inclusion of the S modulus}

We now considers the possible effects of fields stabilized by fluxes in KKLT like models. Indeed, whereas these fields can obtain a large
mass from fluxes which freeze their dynamics at energies below their mass, close to big-bang their dynamics is crucial in order to determine the climbing behavior. We consider for definiteness the axion-dilaton $S$. The corresponding K\"ahler potential and superpotential are
\begin{equation}
 K \ = \ - \ln(S+\bar S) \ - \ 3 \ \ln(T+\bar T) \qquad , \qquad  W \ = \ W_{eff}(S) \ + \ c \ e^{-\tilde bT} \ . \label{kklt7}
\end{equation}
According to \cite{dks} and arguments in Section \ref{seckklt}, we neglect the non-perturbative term $c e^{-\tilde bT}$ close to the Big Bang. The resulting SUGRA scalar potential is of the form
\begin{equation}
V_F \ = \ \frac{|W_{eff}(S)-(S+\bar S) \ W'_{eff}(S)|^2}{(S+\bar S)(T+\bar T)^3} \ .
\label{kklt8}
\end{equation}
In addition we consider un uplift potential given by
\begin{equation}
V_{up} \ = \ \frac{D}{(S+\bar S)^{n_S}(T+\bar T)^{n_T}} \ , \label{kklt9}
\end{equation}
where the uplift parameters $n_S,n_T$ depend on the specific uplift mechanism and are considered, as before, as free parameters in what follows.
We parametrize the complex scalar fields $S$ and $T$ as
\begin{equation}
S \ = \ e^{\Phi_S} + i \theta_S \qquad , \qquad T \ = \ e^{\frac{\Phi_T}{\sqrt{3}}} + i \frac{\theta_T}{\sqrt{3}} \ . \label{kklt10}
\end{equation}
The form of the asymptotic solutions is as follows
\begin{equation}
\Phi_S \ = \ p_S \ln t \quad , \quad \Phi_T \ = \ p_T \ln t \quad , \quad
{\dot \theta}_S \ = \ q_S \ t^{p_S-1} \quad , \quad
{\dot \theta}_T \ = \ q_T \ t^{\frac{p_T}{\sqrt{3}}-1} \ , \label{kklt11}
\end{equation}
where the exponents in the axion fields were chosen such that their kinetic terms behave
as $t^{-2}$ close to big-bang.
We take the effective superpotential for $S$ to be a polynomial of degree $n$
\begin{equation}
W_{eff}(S) \ = \ W_0 \ + \ W_1 \ S \ +...+ \ W_n \ S^n \ . \label{kklt12}
\end{equation}
Indeed, whereas three form fluxes lead originally to linear terms in the dilaton in the superpotential \cite{dudas,gkp}, integrating out other (complex scalar) fields lead to a more general polynomial in $S$. Strictly speaking, according to our previous argument, dynamics of all
fields should be explicitly kept close to big-bang. We however proceed by considering only the axion-dilaton in order to keep the analysis as simple as possible in what follows.
The constants $p_S, p_T, q_S, q_T$ parameterizing the asymptotic solution have then to satisfy the constraints
\begin{equation}
\begin{split}
&p_S^2 + p_T^2 \ = \ \frac{4}{3} \qquad , \qquad q_S \ = \ q_T \ = \ 0 \ , \\
&- p_S \ - \ \sqrt{3} \ p_T \ > \ - 2 \ , \\
&(2n-1) \ p_S \ - \ \sqrt{3} \ p_T \ > \ - 2 \ , \\
&-n_S \ p_S \ - \ \frac{n_T}{\sqrt{3}} \ p_T \ > \ - 2 \ . \\
\end{split} \label{kklt13}
\end{equation}
The three inequalities in (\ref{kklt13}) correspond to regularity conditions on the scalar potential as in eq. (\ref{regularity}), which ensure that the terms involving the potential can be neglected in the limit $t\rightarrow 0$. According to our general analysis in section 5 and specifically the case in Section \ref{stabilizant}, if the following conditions are satisfied
\begin{equation}
n \geq 2 \qquad , \qquad n_S \geq \sqrt{3} \qquad , \qquad n_T \geq 3 \ .
\label{skklt}
\end{equation}
then there is no solution with $p_T <0$, or equivalently with $\dot \Phi_T <0$. This implies that the inequalities in eq. (\ref{skklt}) have to be satisfied in order to obtain a climbing behavior for the scalar $\Phi_T$.
 An immediate consequence of the conditions above is that a superpotential with a linear term in $S$, i.e. with $n=1$ above, would spoil the climbing.

%%%%%%%%%%%%%%%%%%%%%%%%%%%%%%%%%%%%%%%%%%%%%%%%%%%%%%%%%%%%%%%%%%%%%%%%%%%%%%%%%

\section{Conclusions}

An effective field theory analysis shows that, under certain assumptions, a scalar field may be forced to climb its potential when it emerges from the big-bang. This was demonstrated in \cite{dks} by finding the exact solution(s) to a system consisting of a single scalar field propagating in a gravitational background. There are several assumptions involved in this scenario:
(i) all fields, that is the metric and the scalar field, are assumed to have only a dependence on time; (ii) the metric is assumed to have an $SO(d)$ symmetry and thus its dynamics is reduced to that of a scale factor; (iii) there is only one scalar field in the theory subject to an exponential potential $V\sim e^{\lambda\Phi}$ with $\lambda$  above a certain critical value $\lambda_c$ depending on the number of dimensions; (iv) one has to trust the field theory analysis all the way to the big-bang singularity, thus neglecting any possible corrections coming from quantum gravity. Although, as our results point out as well, the existence of a climbing scalar in the very early cosmology is a very special scenario, it is not limited to the case mentioned above. We have shown that climbing can arise in more general scenarios with multiple scalar fields present in the theory. Our analysis relies on solving the equations of motion asymptotically close to the big-bang. This is sufficient in order to hunt for climbing scalars in the very early universe as the climbing phase starts immediately after the big-bang. The strategy is the following. Neglecting the scalar potential allows one to solve the equations of motion for the particular gravitational background mentioned earlier. Generically the solution for the scale factor and for the scalar fields is of Kasner type. Not all solutions found in this way correspond to asymptotic solutions of the full system including the potential. One has to eliminate the solutions which lead to a too singular scalar potential (and its first derivative). Given that the kinetic terms have a singularity of  order $t^{-2}$ it follows that an asymptotic solution $\Phi$ has to satisfy $V(\Phi(t)), \partial_\Phi V(\Phi(t))\sim O(t^{-2+\epsilon})$. The condition for climbing  $\lambda>\lambda_c$, from \cite{dks}, is recovered in this way for the case of one scalar field. Scenarios with more scalar fields have to satisfy more general algebraic constraints in order to have climbing, displayed in Section 6.

An important note is that, although the models considered have more scalars present, only one of them could be forced to climb the potential. In other words, the algebraic constraints necessary in order to have two (or more) scalars forced to climb have no solution. Our analysis was done for the maximal space symmetry  of the background considered in \cite{dks}. The reason is that the climbing phenomenon disappears whenever one relaxes the $SO(d)$ symmetry of the background. We have shown this explicitly for the case of one scalar field. This happens because in the absence of maximal space symmetry, it is possible to find solutions for the scale factors such that the velocity of the scalar field $\Phi$ is arbitrarily small. Thus, for a given exponent $\lambda$ it is always possible to find a descending scalar with a small enough velocity. The bigger $\lambda$ is the smaller is the maximum velocity allowed for a descending solution.

We have also presented exact solutions for the case of multi-exponential scalar potentials, which are generalizations of the Lucchin-Matarrese attractor \cite{lm} to several scalar fields, to be interpreted, in analogy with the one field case, as late-time attractor solutions. Such solutions
are characterized by the fact that all terms in the (exponential) scalar potential behave like
kinetic terms (in $t^{-2}$) at any time. Moreover, we worked out asymptotic solutions for mixed cases, in which some terms behaves like $t^{-2}$ for early times, while other terms are negligible. In all such examples that we analyzed explicitly, all scalar fields are of descending type.

There is an interesting example, studied in Section 4, based on a single scalar field with a potential given by a sum of two exponentials with critical or supercritical exponents and a flipped sign. One term would kill the descending Kasner solution whereas the other would kill the climbing Kasner solution. Thus, for this system we were not able to find any Kasner solution. With the notable exception of the  obvious solution of a de Sitter universe with the scalar field frozen at the minimum of the potential, there are oscillating solutions with scalar oscillations amplifying forever
going backwards in time towards big-bang.

Another direction that we considered in Section 5 is the robustness of the Kasner climbing solution when higher derivative corrections are included. This is relevant if one wants to embed such a climbing scenario in string theory. The correction to the original action contains now curvature square terms and non-minimal couplings of the scalar field to the various components of the Riemann curvature tensor. From a string theory point of view this would correspond to the first $\alpha'$ corrections. Demanding that the new action is ghost free (absence of higher derivatives than two in the equations of motion) and that the Kasner climbing solution for one scalar field be preserved we have deduced a number of conditions on the couplings in the $\alpha'$ corrected action. It turns out that there is essentially an unique action (up to a normalization factor) at squared level in the curvature that is both ghost free and that preserves the climbing. In order to see if the climbing phenomenon could survive the whole series of $\alpha'$ corrections coming from string theory one would need in principle an exactly solvable string background containing a climbing scalar in the effective theory. This is very difficult to achieve in practice.

Finally, we have applied the asymptotic method for hunting climbing scalars to string effective actions with moduli stabilization.  In particular, we considered the KKLT scenario of moduli stabilization. As shown in \cite{dks}, the system is exactly critical for F-terms \cite{fuplift} or D-terms uplift \cite{duplift}, whereas is subcritical for the original antibranes uplift \cite{kklt}.
For KKLT and by keeping only the uplift part of the scalar potential, we find that there is always a descending scalar, but with a fine-tuned initial condition, whereas for other initial conditions the climbing can persist. We also complicated the dynamics by including the axion-dilaton field and deduced a condition on its superpotential in order for the system to still have a climbing for the K\"ahler modulus. As an immediate consequence, we found that the presence of an axion-dilaton with a linear superpotential would spoil the climbing.

The method that we proposed only allows one to decide wether there is or not a climbing scalar in the very early universe. Following the dynamics afterwards requires one to resort either to exact solutions as in our Section 3 or \cite{dks,augusto} or to numerical analysis. In the standard cosmology the climbing phase would be followed by an inflationary epoch. In general, inflation has a tendency to wash away the physical dynamics prior to it. This depends in principle on the number of e-folds. The presence of a climbing phase can lead to observable effects in the later times if the total number of e-folds is not too large. Indeed, recent results in \cite{dkps,oscil,augusto} indicate that a climbing phase would lead to a dampening of the power spectrum on large scales and oscillations superimposed on it that fit relatively well the Planck 2013 data \cite{planck1,planck2,planck3,planck4}.

\begin{figure}[h!]
  \centering
  \includegraphics[width=10cm]{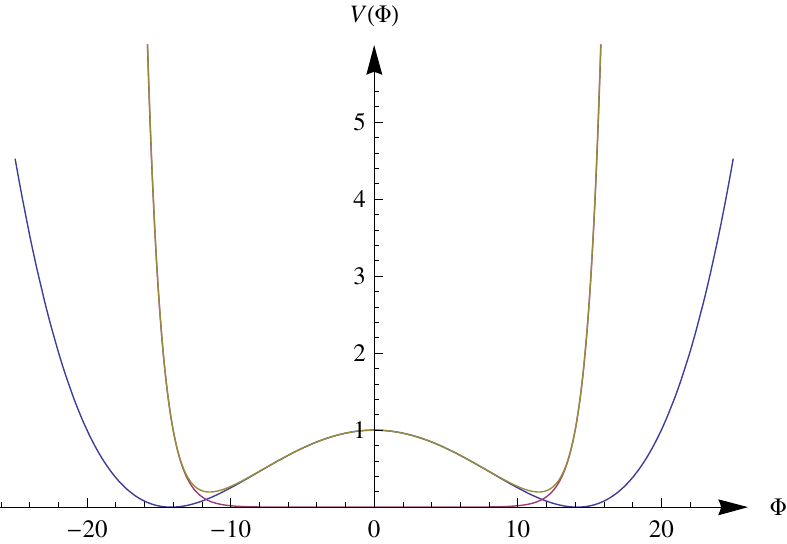}\\
  \caption{Hill-top inflationary potential  $V_1(\Phi)=\left(1-\frac{\Phi}{200}\right)^2$ (blue), abrupt potential as in Section 4 $V_2=e^{\Phi-14}+e^{-\Phi-14}$ (red) and their sum $V_3(\Phi)= V_1(\Phi)+V_2(\Phi)$ (brown). The large field values are inaccessible, the scalar is forced to move within the two abrupt walls where inflation takes place.}\label{inflation}
\end{figure}

Another potential application of the climbing phase of a scalar field is in the issue of initial conditions in inflation. Indeed, it was pointed out recently\footnote{There are however several simple ways to address the problem of
initial conditions in inflation in light of the Planck 2013 data. See
for example the recent talk of A. Linde in KITP,
http://online.kitp.ucsb.edu/online/primocosmo-c13/linde/oh/01.html. We
thank Andrei Linde for correspondence and for pointing out to us the
arguments and the reference above.} that most inflationary potentials favored by the recent Planck 2013 data require somewhat non-generic initial conditions \cite{Ijjas:2013vea}. The climbing potentials we investigated in this paper can force the inflaton to be in the right
parameter space values for successful inflationary predictions. For example, in Figure
\ref{inflation}
we display a hill-top inflationary potential (see e.g. \cite{olive})  $V = \Lambda^4
(1-\Phi^2/\mu^2)^2$, with $\mu >> M_P$. Planck 2013 prefers the flat region
$|\Phi| < \mu$, whereas most of the region in field space, for $\Phi > \mu$, corresponds
to a $\Phi^4$ potential which is highly disfavored by the Planck data. By adding on top of this Higgs-like potential a steep  potential
$V \sim \alpha (e^{\lambda (\Phi-\Phi_0)} + e^{\lambda (-\Phi-\Phi_0)})$, with
$\Phi_0 \le \mu$, similarly to the example discussed in Section 4, the scalar field
is forced to stay  within the field region
$|\Phi| < \Phi_0 < \mu$ most of the time,  whereas in the short initial times when it goes outside this region it cannot inflate, improving therefore the initial value problems for inflation. Lastly, it would be interesting to investigate the effect of the climbing dynamics on the chaotic BKL behavior \cite{bkl} in general relativity.

%\begin{figure}[h!]
%  \centering
%  % Requires \usepackage{graphicx}
%  \includegraphics[width=10cm]{inflation}\\
%  \caption{The following functions are represented $V_1(\Phi)=\left(1-\frac{\Phi}{200}\right)^2$, %$V_2=e^{\Phi-10}+e^{-\Phi-10}$ and $V_3(\Phi)= V_1(\Phi)+V_2(\Phi)$.}\label{inflation}
%\end{figure}

%\begin{figure}[h!]
%  \centering
%  % Requires \usepackage{graphicx}
%  \includegraphics[width=10cm]{inflationter}\\
%  \caption{The following functions are represented $V_1(\Phi)=\left(1-\frac{\Phi}{200}\right)^2$, %$V_2=e^{2(\Phi-14)}+e^{-2(\Phi+14)}$ and $V_3(\Phi)= V_1(\Phi)+V_2(\Phi)$.}\label{inflation}
%\end{figure}

%%%%%%%%%%%%%%%%%%%%%%%%%%%%%%%%%%%%%%%%%%%%%%%%%%%%%%%%%%%%%%%%%%%%%%%%%%%%%%%%%%%%%%%%%
\

\noindent {\bf Acknowledgements. }  We would like  to thank Jihad Mourad for discussions and
especially Augusto Sagnotti for enlightening discussions throughout the duration of the project and collaboration in Section 4 of our paper. C.C. is very grateful to Ecole Polytechnique (CPHT) and to CERN Theory Unit for hospitality during various stages of this work. E.D. thanks the Galileo Galilei Institute for Theoretical Physics for the hospitality and the INFN for partial support during the final part of the completion of this work. This  work was supported in part by the European ERC Advanced Grant 226371 MassTeV, the French ANR TAPDMS ANR-09-JCJC-0146, the contract PITN-GA-2009-237920 UNILHC and the CNCS-UEFISCDI grant PD 103/2012.

%%%%%%%%%%%%%%%%%%%%%%%%%%%%%%%%%%%%%%%%%%%%%%%%%%%%%%%%%%%%%%%%%%%%%%%%%%%%%%%%%%%%%%%%%%%%%%%%%%%%%%%%%%%%%%%%%%%%%%%%%%%%%%%%
\appendix

\section{Higher-derivative corrections}

One way of deriving the equations of motion is to write the action $S_1+S_2$ in terms of $A(t)$ and $\Phi(t)$ and then taking the variation with respect to these. There are, however,  three independent equations, two Einstein equations and one corresponding to the scalar field $\Phi$. In order to get all the equations it is necessary to introduce a lapse function $B$:
\begin{equation}
ds^2 = -e^{2B(t)}dt^2+e^{2A(t)}\sum_{i=1}^d(dx^i)^2
\label{metricB}
\end{equation}
Then taking the variation of the action with respect to $A(t), B(t)$ and $\Phi(t)$ yield the three equations of motion. At the end we shall impose the condition
\begin{equation}
B=0
\end{equation}

For the metric tensor in eq. (\ref{metricB}) one obtains the following expressions for the Riemann tensor $R_{\mu\nu\rho\sigma}$, Ricci tensor $R_{\mu\nu}$ and curvature scalar $R$:
\begin{equation}
R_{00}=-d(\ddot A +\dot A^2-\dot A\dot B)  \qquad R_{ij}=e^{2A-2B}(\ddot A +d\dot A ^2-\dot A\dot B)\delta_{ij}
\end{equation}
\begin{equation}
R_{0i0j}=-e^{2A}(\ddot A+\dot A^2-\dot A\dot B)\delta_{ij} \qquad R_{ijkl}=e^{4A-2B}\dot A^2(\delta_{ik}\delta_{jl}-\delta_{il}\delta_{jk})
\end{equation}
\begin{equation}
R=e^{-2B}(2d\ddot A +d(d+1)\dot A^2-2d \dot A\dot B)
\end{equation}
where the indices $i,j,k,l=1,...,d$.
%\begin{equation}
%\begin{split}
%&R_{0i0k} \ = \ - e^{2A} \left(\ddot A+\dot A^2 \right) \delta_{ik} \ , \\
%&R_{00} \ = \ - d \left(\ddot A+\dot A^2\right) \ , \\
%\end{split}
%\qquad
%\begin{split}
%&R_{ijkl} \ = \ e^{4A} \dot A^2\left(\delta_{ik}\delta_{jl}-\delta_{il}\delta_{jk}\right) \ , \\
%&R_{ij} \ = \ e^{2A} \left(\ddot A +d \dot A^2\right)\delta_{ij} \ ,
%\end{split}
%\end{equation}
%\begin{equation}
%R \ = \ 2d \ddot A  \ + \ d(d+1) \dot A^2 \ ,
%\end{equation}

It is useful to define the following combinations of coefficients
\begin{equation}
\begin{split}
&c_1 \ = \ 4\alpha_1d+\alpha_2(d+1)+4\alpha_3 \ , \\
&c_2 \ = \ \alpha_1d(d+1)+\alpha_2 d+2\alpha_3\ . \\
\end{split}
\end{equation}
Making use of the equations above one can easily show that the actions $S_1$ and $S_2$ can be rewritten as
\begin{align}
S_1&=\frac{1}{2}\int d^{d+1}x\, e^{dA-3B-q\Phi}d\left[c_1 \ddot A^2 +4c_2 \ddot A \dot A^2 +(d+1)c_2 \dot A^4+c_1 (\dot A^2\dot B^2-2\dot A\dot B \ddot A) -4c_2 \dot A^3 \dot B\right]\ , \\
S_2&=\frac{1}{2}\int d^{d+1} x e^{dA-3B-q\Phi}\left\{\left[2d\ddot A +d(d+1)\dot A^2-2d \dot A\dot B\right]\times\left[-\beta_1\dot\Phi^2-\beta_6
(\ddot\Phi+d\dot A\dot\Phi-\dot B\dot\Phi)\right]\right.\nonumber\\
&+d(\ddot A+\dot A^2 -\dot A\dot B)\times\left[-\beta_2\dot\Phi^2-\beta_7(\ddot\Phi+\dot B\dot\Phi)\right]+d(\ddot A +d \dot A^2-\dot A\dot B)\beta_7 \dot A \dot\Phi+\beta_3 \dot\Phi^4\nonumber\\
&\left.+(\ddot\Phi+d\dot A\dot \Phi-\dot B\dot\Phi)\times\left[\beta_4(\ddot\Phi+d\dot A\dot\Phi-\dot B\dot\Phi)+\beta_5\dot\Phi^2\right]\right\}
\end{align}
%Notice that the variation of $S_1$ contributes only to the Einstein equation of motion for $A(t)$.
By taking the variation the variation of the action $S_1+S_2$ with respect to $A(t), B(t)$ and $\Phi$ and demanding the absence of derivatives higher than two in the equations of motion one arrives at the following conditions
\begin{eqnarray}
c_2=0\\
2\beta_1+\beta_2=0\\
\beta_4=\beta_6=\beta_7=0
\end{eqnarray}
Notice that the condition $c_2=0$ is satisfied by the Gauss-Bonnet term, that is for the following choice of parameters
\begin{equation}
\alpha_1=\eta  \qquad \alpha_2=-4\eta \qquad \alpha_3=\eta
\end{equation}
with $\eta$ a real arbitrary parameter. In fact, this is the unique combination (up to an overall normalization $\eta$) which does not contain higher derivatives in the equation of motion (see \cite{Zwiebach:1985uq}). It does not follow from our calculation due to the fact that we consider pure time dependent solutions with $SO(d)$ symmetry.
Imposing the conditions of absence of higher derivative terms leads one to consider the following action
\begin{align}
S & =\frac{1}{2}\int d^{d+1}x \sqrt{-g}e^{-q\Phi}\Big[\eta \left(R^2-4R_{\mu\nu}R^{\mu\nu}+R_{\mu\nu\rho\sigma}R^{\mu\nu\rho\sigma}\right)\nonumber \\
 &-2\beta_1 G^{\mu\nu}\nabla_\mu \Phi\nabla_\nu\Phi+\beta_3 (\nabla\Phi)^4+\beta_5 (\nabla\Phi)^2\Box\Phi\Big]
\end{align}
Explicitly, in terms of $A(t), B(t)$ and $\Phi(t)$ one has
\begin{align}
S_1&= \frac{1}{2}\int d^{d+1}x e^{dA-3B-q\Phi}\, d(d-1)(d-2)\eta\left[4 \ddot A \dot A^2 +(d+1)\dot A^4 -4\dot A^3 \dot B\right]\\
S_2&=\frac{1}{2}\int d^{d+1}x e^{dA-3B-q\Phi}\left[-d(d-1)\beta_1 \dot A^2+\beta_3\dot\Phi^2+\beta_5(\ddot \Phi+d\dot A\dot\Phi-\dot B\dot\Phi)\right]\dot\Phi^2
\end{align}
The equations of motion, after putting $B=0$, are then found to be:
\begin{align}
\delta_B S&=\frac{1}{2}e^{dA-q\Phi}\Big\{d (d-1)(d-2)\eta[(d-3)\dot A^4 -4q \dot A^3 \dot\Phi] +3d(d-1)\beta_1 \dot A^2\dot\Phi^2 \nonumber\\
& - (3\beta_3+q\beta_5)\dot\Phi^4 -2d \beta_5 \dot A\dot\Phi^3\Big\}\\
\delta_A S &=\frac{1}{2}e^{dA-q\Phi}\Big\{d(d-1)(d-2)\eta\Big[4(d-3)\dot A^2 \ddot A +d(d-3)\dot A^4 +4q^2 \dot A^2 \dot\Phi^2-4q(d-1)\dot A^3 \dot\Phi\nonumber\\
&-8q\dot A\ddot A \dot\Phi -4q \dot A^2\ddot\Phi\Big]+d\Big[d(d-1)\beta_1\dot A^2\dot\Phi^2
-2q(d-1)\beta_1\dot A\dot\Phi^3+(\beta_3+q\beta_5)\dot\Phi^4-2\beta_5\dot\Phi^2\ddot\Phi\nonumber\\
&+2(d-1)\beta_1\ddot A\dot\Phi^2+8(d-1)\beta_1\dot A\dot\Phi\ddot\Phi\Big]\Big\}\\
\delta_\Phi S&=\frac{1}{2}e^{dA-q\Phi}\Big\{-q(d-1)(d-2)\eta\left[4d\dot A^2\ddot A+d(d+1)\dot A^4\right]-2d\beta_5 \ddot A\dot\Phi^2\nonumber\\
&-d[(d-1)q\beta_1+2d\beta_5]
\dot A^2\dot\Phi^2 +2d(d-1)\beta_1 \dot A\ddot A \dot\Phi+2d^2(d-1)\beta_1\dot A^3\dot\Phi\nonumber\\
&+d(d-1)\beta_1\dot A^2\ddot\Phi-4d\beta_3 \dot A\dot\Phi^3-4d\beta_5\dot A\dot\Phi\ddot\Phi+q(3\beta_3+q\beta_5)\dot\Phi^4-4(3\beta_3+q\beta_5)\dot\Phi^2\ddot\Phi\Big\}
\end{align}

Finally, we demand that the corrections of order $t^{-4}$ preserve the (asymptotic) solution in eq. (\ref{solutions}). This amounts to imposing that we must have
\begin{equation}\label{action variation}
\delta_A S\ = \ \delta_B S\ =\ \delta_\Phi S \ = \ 0
\end{equation}
for the respective solutions. Plugging in $A=\frac{1}{d}\ln t$ and $\Phi=p\ln t$ one can show that the coefficients $\beta_1, \beta_3, \beta_5$ and $\eta$ have to satisfy the following conditions
\begin{eqnarray}
\beta_1&=&0 \label{q1} \label{c1}\\
\beta_3&=& \frac{(d-2)\left[-3(d-3)+3q^2 dp^2+q(d+3)p\right]}{12d(d-1)}\eta\label{q2}\label{c2}\\
\beta_5&=&\frac{(d-2)\left[2(d-3)-3q^2dp^2-q(5d+3)p\right]p}{2d(d-1)(qp+2)}\eta \label{c3}
\end{eqnarray}
We conclude that there is a unique (up to normalization) combination of terms of order $t^{-4}$ preserving the asymptotic solution in eq. (\ref{solutions}). Consider, for instance, the simple case of $q=0$, then the eqs. (\ref{c1}), (\ref{c2}), (\ref{c3}) simplify to
 \begin{equation}
\beta_1=0 \qquad \beta_3 =-\frac{(d-2)(d-3)}{4d(d-1)}\eta \qquad \beta_5=\frac{(d-2)(d-3)p}{2d(d-1)}\eta
\end{equation}
and the corresponding action preserving the asymptotic solutions in eq. (\ref{solutions}) is given by
\begin{align}
S& =\frac{1}{2}\int d^{d+1}x \sqrt{-g}\eta \Bigg\{R^2-4R_{\mu\nu}R^{\mu\nu}+R_{\mu\nu\rho\sigma}R^{\mu\nu\rho\sigma}\nonumber\\
&-\frac{(d-2)(d-3)}{4d(d-1)}\left[(\nabla\Phi)^4-2 \sqrt{\frac{2(d-1)}{d}}(\nabla\Phi)^2\Box\Phi\right]\Bigg\}
\end{align}
Finally, in the general case, for $q$ arbitrary we have that the following action preserves the solutions in eq. (\ref{solutions})
\begin{equation}
\begin{split}
S&=\frac{1}{2}\int d^{d+1}x \sqrt{-g}\,\eta\, e^{-q\Phi}\Bigg\{R^2-4R_{\mu\nu}R^{\mu\nu}+R_{\mu\nu\rho\sigma}R^{\mu\nu\rho\sigma}\\
&-\frac{(d-2)}{12d(d-1)}\Big\{\left[3(d-3)-3q^2dp^2-q(d+3)p\right](\nabla\Phi)^4\\
&-6\left[2(d-3)-3q^2dp^2-q(5d+3)p\right] \frac{p}{qp+2}(\nabla\Phi)^2\Box\Phi\Big\}\Bigg\}
\end{split}
\end{equation}

%%%%%%%%%%%%%%%%%%%%%%%%%%%%%%%%%%%%%%%%%%%%%%%%%%%%%%%%%%%%%%%%%%%%%%%%%%%%%%%%%%%%%%%%%%%%%%%%%%%
\section{String frame vs Einstein frame}
\label{string frame}
In this appendix we present formulas for relating the parameters in the string frame to the ones in the Einstein frame. We define the following action in the string frame
\begin{equation}
\begin{split}
S&=\int d^{d+1}x\sqrt{-\tilde g}\left[e^{-2\varphi}\left( \tilde R +4(\tilde\nabla\varphi)^2\right)+e^{-\tilde q\varphi}\left(\tilde\alpha_1\tilde R^2+\tilde\alpha_2\tilde R_{\mu\nu}\tilde R^{\mu\nu}+\tilde\alpha_3\tilde R_{\mu\nu\rho\sigma}\tilde R^{\mu\nu\rho\sigma}\right)\right.\\
&+\left.e^{-\tilde q\varphi}\left(\tilde\beta_1 \tilde R (\tilde \nabla \varphi)^2+\tilde\beta_2 \tilde R^{\mu\nu}\tilde\nabla_\mu\varphi\tilde\nabla_\nu\varphi+\tilde\beta_3 (\tilde\nabla\varphi)^4
+\tilde\beta_4 (\tilde\Box\varphi)^2+\tilde\beta_5 (\tilde\nabla\varphi)^2 \tilde\Box\varphi\right.\right.\\
&\left.\left.+\tilde\beta_6\tilde R\tilde\Box\varphi+\tilde\beta_7 \tilde R^{\mu\nu}\tilde\nabla_\mu\tilde\nabla_\nu\varphi\right)\right]
\end{split}
\end{equation}
The transformation to the Einstein frame is defined by a Weyl rescaling of the metric tensor
\begin{equation}
\tilde g _{\alpha\beta} =e^{2\omega}g_{\alpha\beta}
\label{weyl}
\end{equation}
with the scalar function $\omega$ given by
\begin{equation}
\omega=\frac{2}{d-1}\varphi
\end{equation}

One further needs to do the following redefinition of the scalar field $\varphi$ in order to obtain a canonically normalized kinetic term for the field $\Phi$:
\begin{equation}
\varphi=\sqrt{\frac{d-1}{8}}\Phi
\end{equation}
Making use of the transformation rules of the Ricci scalar, Ricci tensor and Riemann tensor under the Weyl rescaling in eq. (\ref{weyl})
\begin{align}
\tilde R &= e^{-2\omega}\left[R-2d\,\Box\omega-d(d-1)(\nabla\omega)^2\right]\\
\tilde R_{\mu\nu}&= R_{\mu\nu} - (d-1)\nabla_\mu\nabla_\nu \omega -g_{\mu\nu}\Box\omega+(d-1) \nabla_\mu\omega\nabla_\nu\omega-(d-1)g_{\mu\nu}(\nabla\omega)^2\\
\tilde{R}_{\mu\nu\rho\sigma}&= e^{2\omega}\left[R_{\mu\nu\rho\sigma}+2g_{\sigma[\mu}\nabla_{\nu]}\nabla_\rho\omega-2g_{\rho[\mu}\nabla_{\nu]}\nabla_\sigma\omega
+2\nabla_{[\mu}\omega g_{\nu]\sigma}\nabla_\rho\omega\right.\nonumber\\
&\left. -2\nabla_{[\mu}\omega g_{\nu]\rho}\nabla_\sigma\omega-2g_{\rho[\mu}g_{\nu]\sigma}(\nabla \omega)^2\right]
\end{align}
one can show that the resulting action in the Einstein frame is given by
\begin{equation}
\begin{split}
S_E &=\int d^{d+1}x\sqrt{- g}\left[\left( R -\frac{1}{2}(\nabla\Phi)^2\right)+e^{- q \Phi}\left(\alpha_1 R^2+\alpha_2 R_{\mu\nu} R^{\mu\nu}+\alpha_3 R_{\mu\nu\rho\sigma} R^{\mu\nu\rho\sigma}\right)\right.\\
&+\left.e^{-q\Phi}\left(\beta_1  R ( \nabla \Phi)^2+\beta_2 R^{\mu\nu}\nabla_\mu\Phi\nabla_\nu\Phi+\beta_3 (\nabla\Phi)^4
+\beta_4 (\Box\Phi)^2+\beta_5 (\nabla\Phi)^2 \Box\Phi\right.\right.\\
&\left.\left.+\beta_6 R\Box\Phi+\beta_7  R^{\mu\nu}\nabla_\mu\nabla_\nu\Phi\right)\right]
\end{split}
\end{equation}
were the parameters in the Einstein frame $\{\alpha_i\}_{i=1,2,3}$ and $ \{\beta_j\}_{j=1,...,7}$ are related to the ones in the string frame by
\begin{equation}
\alpha_1=\tilde \alpha_1 \qquad \alpha_2=\tilde \alpha_2 \qquad \alpha_3=\tilde \alpha_3
\end{equation}
\begin{align}
\beta_1&=-d\tilde \alpha_1 -\tilde \alpha_2 -\frac{2}{d-1}\tilde \alpha_3+\frac{d-1}{8}\left(\tilde \beta_1+2\tilde \beta_6\right)+\frac{1}{4}\tilde\beta_7\\
\beta_2&=-\frac{d-3}{2}\tilde\alpha_2-2\frac{d-2}{d-1}\tilde\alpha_3+\frac{d-1}{8}\tilde \beta_2-\frac{d-3}{4(d-1)}\tilde\beta_7\\
\beta_3&=\left(\frac{d-1}{8}\right)^2\Bigg\{ 16\left(\frac{d}{d-1}\right)^2\tilde\alpha_1+2\left[q'^2+4\frac{2d-q'(d-1)}{d-1}\right]\tilde\alpha_2\\
&+\frac{16}{(d-1)^2}\left[\frac{1}{2}q'^2d+\frac{2d-q'(2d-1)}{d-1}\right]\tilde\alpha_3-4\frac{d}{d-1}\tilde \beta_1-q' \tilde \beta_2+\tilde \beta_3\nonumber\\
&+4\tilde\beta_4 +2\tilde\beta_5-8\frac{d}{d-1}\tilde\beta_6-\left[q'^2+\frac{8d-6q'(d-1)}{(d-1)^2}\right]\tilde\beta_7\Bigg\}\nonumber\\
\beta_4&=  \frac{1}{2(d-1)}\Big[4d^2\tilde\alpha_1+d(d+1)\tilde\alpha_2+4(d+1)\tilde\alpha_3\Big]+\frac{d-1}{8}\tilde\beta_4-\frac{d}{4}
\left(2\tilde\beta_6+\tilde \beta_7\right)\\
\beta_5 &=\left(\frac{d-1}{8}\right)^{3/2}\Bigg\{32 \left(\frac{d}{d-1}\right)^2\tilde\alpha_1+\left[-6q'+8\frac{5d-3}{(d-1)^2}\right]\tilde\alpha_2 +\frac{16}{(d-1)^2}\left[\frac{6d-5}{d-1}-\frac{3}{2}q'd\right]\tilde\alpha_3\nonumber\\
&-4\frac{d}{d-1}\tilde \beta_1+\frac{d-3}{d-1}\tilde\beta_2+4\tilde\beta_4+\tilde\beta_5-12\frac{d}{d-1}\tilde\beta_6+\left(3q'-\frac{18}{d-1}\right)
\tilde\beta_7\Bigg\}\\
\beta_6&=\sqrt{\frac{d-1}{8}}\left[-\frac{4}{d-1}\Big(2d\tilde\alpha_1+\tilde\alpha_2\Big)+\tilde\beta_6\right]\\
\beta_7&=\sqrt{\frac{d-1}{8}}\left(-4\tilde\alpha_2-\frac{16}{d-1}\tilde\alpha_3+\tilde\beta_7\right)
\end{align}
Finally the power of dilaton in the Einstein frame $q$ is related to the one in the string frame $\tilde q$ by the following relation
\begin{align}
q=\left(\frac{d-1}{8}\right)^{1/2}q'=\left(\frac{d-1}{8}\right)^{1/2}\left(\tilde q -2\frac{d-3}{d-1}\right)\ .
\end{align}

%%%%%%%%%%%%%%%%%%%%%%%%%%%%%%%%%%%%%%%%%%%%%%%%%%%%%%%%%%%%%%%%%%%%%%%%%%%%%%%%%%%%%%%%%%%%%%%%%%
%%%%%%%%%%%%%%%%%%%%%%%%%%%%%%%%%%%%%%%%%%%%%%%%%%%%%%%%%%%%%%%%%%%%%%%%%%%%%%%%%%%%%%%%%%%%%%%%%%%

%%%%%%%%%%%%%%%%%%%%%%%%%%%%%%%%%%%%%%%%%%%%%%%%%%%%%%%%%%%%%%%%%%%%%%%%%%%%%%%%%%%%%%%%%%%%%%%%%%%%%%%%%%%%%%%%%%%%%%%%%%%%%%%


\begin{thebibliography}{99}


\bibitem{inflation1}
 A.~A.~Starobinsky,
  ``A New Type of Isotropic Cosmological Models Without Singularity,''
  Phys.\ Lett.\ B {\bf 91} (1980) 99;
  %%CITATION = PHLTA,B91,99;%%
 A.~H.~Guth,
  ``The Inflationary Universe: A Possible Solution to the Horizon and Flatness Problems,''
  Phys.\ Rev.\ D {\bf 23} (1981) 347;
  %%CITATION = PHRVA,D23,347;%%
 A.~D.~Linde,
  ``A New Inflationary Universe Scenario: A Possible Solution of the Horizon, Flatness, Homogeneity, Isotropy and Primordial Monopole Problems,''
  Phys.\ Lett.\ B {\bf 108} (1982) 389;
  %%CITATION = PHLTA,B108,389;%%
   A.~Albrecht and P.~J.~Steinhardt,
  ``Cosmology for Grand Unified Theories with Radiatively Induced Symmetry Breaking,''
  Phys.\ Rev.\ Lett.\  {\bf 48} (1982) 1220;
   A.~D.~Linde,
  ``Chaotic Inflation,''
  Phys.\ Lett.\ B {\bf 129} (1983) 177;
  %%CITATION = PHLTA,B129,177;%%
For recent reviews see:\\
V.~Mukhanov, ``Physical foundations of Cosmology,'' {\it  Cambridge, UK: Univ.
Pr. (2005)}; \\ S.~Weinberg, ``Cosmology,'' {\it Oxford, UK: Oxford Univ. Pr.
(2008)}.

\bibitem{halliwell}
 J.~J.~Halliwell,
  ``Scalar Fields in Cosmology with an Exponential Potential,''
  Phys.\ Lett.\ B {\bf 185} (1987) 341;
  %%CITATION = PHLTA,B185,341;%%
J.~M.~Aguirregabiria, A.~Feinstein and J.~Ibanez,
  ``Exponential potential scalar field universes. 1. The Bianchi I models,''
  Phys.\ Rev.\ D {\bf 48} (1993) 4662
  [gr-qc/9309013];
  J.~M.~Aguirregabiria, A.~Feinstein and J.~Ibanez,
  ``Exponential potential scalar field universes. 2. The Inhomogeneous models,''
  Phys.\ Rev.\ D {\bf 48}, 4669 (1993)
  [gr-qc/9309014].

\bibitem{dm}
E.~Dudas and J.~Mourad,
  ``Brane solutions in strings with broken supersymmetry and dilaton tadpoles,''
  Phys.\ Lett.\ B {\bf 486} (2000) 172
  [hep-th/0004165].
  %%CITATION = HEP-TH/0004165;%%


\bibitem{townsend}
P.~K.~Townsend and M.~N.~R.~Wohlfarth,
  ``Accelerating cosmologies from compactification,''
  Phys.\ Rev.\ Lett.\  {\bf 91} (2003) 061302
  [arXiv:hep-th/0303097];
  %%CITATION = PRLTA,91,061302;%%
 Class.\ Quant.\ Grav.\  {\bf 21} (2004) 5375
  [arXiv:hep-th/0404241];
  %%CITATION = CQGRD,21,5375;%%
  N.~Ohta,
  ``Accelerating Cosmologies from S-Branes,''
  Phys.\ Rev.\ Lett.\  {\bf 91} (2003) 061303
  [arXiv:hep-th/0303238];
  %%CITATION = PRLTA,91,061303;%%
S.~Roy,
  ``Accelerating cosmologies from M/string theory compactifications,''
  Phys.\ Lett.\  B {\bf 567} (2003) 322
  [arXiv:hep-th/0304084];
  %%CITATION = PHLTA,B567,322;%%
R.~Emparan and J.~Garriga,
  ``A note on accelerating cosmologies from compactifications and S-branes,''
  JHEP {\bf 0305} (2003) 028
  [arXiv:hep-th/0304124];
  %%CITATION = JHEPA,0305,028;%%
   E.~Bergshoeff, A.~Collinucci, U.~Gran, M.~Nielsen and D.~Roest,
  ``Transient quintessence from group manifold reductions or how all roads lead to Rome,''
  Class.\ Quant.\ Grav.\  {\bf 21} (2004) 1947
  [hep-th/0312102];
  %%CITATION = HEP-TH/0312102;%%
J.~G.~Russo,
  ``Exact solution of scalar-tensor cosmology with exponential potentials  and
  transient acceleration,''
  Phys.\ Lett.\  B {\bf 600} (2004) 185
  [arXiv:hep-th/0403010].
  %%CITATION = PHLTA,B600,185;%%


\bibitem{dks}
  E.~Dudas, N.~Kitazawa, A.~Sagnotti,
  ``On Climbing Scalars in String Theory",
  Phys.\ Lett.\  {\bf B694}, 80-88 (2010).
  [arXiv:1009.0874 [hep-th]].


\bibitem{dkps}
 E.~Dudas, N.~Kitazawa, S.~P.~Patil and A.~Sagnotti,
  ``CMB Imprints of a Pre-Inflationary Climbing Phase,''
  arXiv:1202.6630 [hep-th].
  %%CITATION = ARXIV:1202.6630;%%

\bibitem{orientifolds}
A.~Sagnotti, in Cargese '87, ``Non-Perturbative Quantum Field
Theory'', eds. G. Mack et al (Pergamon Press, 1988), p. 521,
%``Open Strings And Their Symmetry Groups,''
arXiv:hep-th/0208020;
%%CITATION = HEP-TH 0208020;%%
G.~Pradisi and A.~Sagnotti,
%``Open String Orbifolds,''
Phys.\ Lett.\ B {\bf 216} (1989) 59;
%%CITATION = PHLTA,B216,59;%%
P.~Horava,
%``Strings On World Sheet Orbifolds,''
Nucl.\ Phys.\ B {\bf 327} (1989) 461,
%%CITATION = NUPHA,B327,461;%%
%``Background Duality Of Open String Models,''
Phys.\ Lett.\ B {\bf 231} (1989) 251;
%%CITATION = PHLTA,B231,251;%%
M.~Bianchi and A.~Sagnotti,
%``On The Systematics Of Open String Theories,''
Phys.\ Lett.\ B {\bf 247} (1990) 517,
%%CITATION = PHLTA,B247,517;%%
%``Twist Symmetry And Open String Wilson Lines,''
Nucl.\ Phys.\ B {\bf 361} (1991) 519;
%%CITATION = NUPHA,B361,519;%%
M.~Bianchi, G.~Pradisi and A.~Sagnotti,
%``Toroidal compactification and symmetry breaking in open string theories,''
Nucl.\ Phys.\ B {\bf 376} (1992) 365;
%%CITATION = NUPHA,B376,365;%%
A.~Sagnotti,
 %``A Note on the Green-Schwarz mechanism in open string theories,''
 Phys.\ Lett.\  B {\bf 294}, 196 (1992)
 [arXiv:hep-th/9210127]~;
 %%CITATION = PHLTA,B294,196;%%
\\
For reviews see: E.~Dudas,
%``Theory and phenomenology of type I strings and M-theory,''
Class.\ Quant.\ Grav.\  {\bf 17}, (2000) R41 [arXiv:hep-ph/0006190];
%%CITATION = HEP-PH 0006190;%%
C.~Angelantonj and A.~Sagnotti,
%``Open strings,''
Phys.\ Rept.\  {\bf 371} (2002) 1 [Erratum-ibid.\  {\bf 376} (2003)
339] [arXiv:hep-th/0204089];
%%CITATION = HEP-TH 0204089;%%
R.~Blumenhagen, B.~Kors, D.~Lust and S.~Stieberger,
 %``Four-dimensional String Compactifications with D-Branes, Orientifolds   and
 %Fluxes,''
 Phys.\ Rept.\  {\bf 445} (2007) 1
 [arXiv:hep-th/0610327].
 %%CITATION = PRPLC,445,1;%%



\bibitem{bsb}
S.~Sugimoto,
``Anomaly cancellations in type I D9-D9-bar system and the USp(32)  string
theory,'' Prog.\ Theor.\ Phys.\  {\bf 102} (1999) 685 [arXiv:hep-th/9905159];
%%CITATION = HEP-TH 9905159;%%
I.~Antoniadis, E.~Dudas and A.~Sagnotti,
``Brane supersymmetry breaking,''
Phys.\ Lett.\ B {\bf 464} (1999) 38 [arXiv:hep-th/9908023];
%%CITATION = HEP-TH 9908023;%%
C.~Angelantonj,
``Comments on open-string orbifolds with a non-vanishing B(ab),''
Nucl.\ Phys.\ B {\bf 566} (2000) 126 [arXiv:hep-th/9908064];
%%CITATION = HEP-TH 9908064;%%
G.~Aldazabal and A.~M.~Uranga,
``Tachyon-free non-supersymmetric type IIB orientifolds via  brane-antibrane
%systems,''
JHEP {\bf 9910} (1999) 024 [arXiv:hep-th/9908072];
%%CITATION = HEP-TH 9908072;%%
C.~Angelantonj, I.~Antoniadis, G.~D'Appollonio, E.~Dudas and
A.~Sagnotti,
``Type I vacua with brane supersymmetry breaking,''
Nucl.\ Phys.\ B {\bf 572} (2000) 36 [arXiv:hep-th/9911081].
%%CITATION = HEP-TH 9911081;%%


\bibitem{kklt}
S.~Kachru, R.~Kallosh, A.~Linde and S.~P.~Trivedi,
  ``De Sitter vacua in string theory,''
  Phys.\ Rev.\  D {\bf 68} (2003) 046005
  [arXiv:hep-th/0301240].

\bibitem{Penrose:1999vj}
  R.~Penrose,
  ``The question of cosmic censorship,''
  J.\ Astrophys.\ Astron.\  {\bf 20}, 233 (1999).

\bibitem{Virbhadra:2002ju}
  K.~S.~Virbhadra and G.~F.~R.~Ellis,
  ``Gravitational lensing by naked singularities,''
  Phys.\ Rev.\ D {\bf 65}, 103004 (2002);
  K.~S.~Virbhadra and C.~R.~Keeton,
  ``Time delay and magnification centroid due to gravitational lensing by black holes and naked singularities,''
  Phys.\ Rev.\ D {\bf 77}, 124014 (2008)
  [arXiv:0710.2333 [gr-qc]].

\bibitem{augusto}
 A.~Sagnotti,
  ``Brane SUSY Breaking and Inflation: Implications for Scalar Fields and CMB Distorsion,''
  arXiv:1303.6685 [hep-th];
  %%CITATION = ARXIV:1303.6685;%%
P.~Fre, A.~Sagnotti, A.S.~Sorin, to appear.


%\cite{Gouteraux:2011qh}
\bibitem{Gouteraux1}
  M.~Cvetic,
``Extreme domain wall - black hole complementarity in N=1 supergravity
with a general dilaton coupling,''
Phys.\ Lett.\ B {\bf 341}, 160 (1994) [hep-th/9402089];
   %%CITATION = HEP-TH/9402089;%%
 M.~Cvetic and H.~H.~Soleng,
  ``Naked singularities in dilatonic domain wall space times,''
   Phys.\ Rev.\ D {\bf 51}, 5768 (1995)
   [hep-th/9411170];
   C.~Martinez, R.~Troncoso and J.~Zanelli,
  ``Exact black hole solution with a minimally coupled scalar field,''
  Phys.\ Rev.\ D {\bf 70} (2004) 084035
  [hep-th/0406111].
  %%CITATION = HEP-TH/0406111;%%
  B.~Gouteraux, J.~Smolic, M.~Smolic, K.~Skenderis and M.~Taylor,
  ``Holography for Einstein-Maxwell-dilaton theories from generalized dimensional reduction,''
  JHEP {\bf 1201}, 089 (2012)
  [arXiv:1110.2320 [hep-th]];
  %%CITATION = ARXIV:1110.2320;%%
 M.~Cadoni, S.~Mignemi and M.~Serra,
  ``Black brane solutions and their solitonic extremal limit in Einstein-scalar gravity,''
  Phys.\ Rev.\ D {\bf 85} (2012) 086001
  [arXiv:1111.6581 [hep-th]].
  %%CITATION = ARXIV:1111.6581;%%




\bibitem{lm}
  F.~Lucchin and S.~Matarrese,
  ``Power Law Inflation,''
  Phys.\ Rev.\ D {\bf 32} (1985) 1316;
  %%CITATION = PHRVA,D32,1316;%%
L.~F.~Abbott and M.~B.~Wise,
 ``Constraints On Generalized Inflationary Cosmologies,''
 Nucl.\ Phys.\  B {\bf 244} (1984) 541;
 %%CITATION = NUPHA,B244,541~;%%
D.~H.~Lyth and E.~D.~Stewart,
 ``The Curvature perturbation in power law (e.g. extended)
 inflation,''
 Phys.\ Lett.\  B {\bf 274} (1992) 168;
 %%CITATION = PHLTA,B274,168~;%%
I.~P.~C.~Heard and D.~Wands,
  ``Cosmology with positive and negative exponential potentials,''
  Class.\ Quant.\ Grav.\  {\bf 19} (2002) 5435
  [arXiv:gr-qc/0206085];
  %%CITATION = CQGRD,19,5435;%%
 ``The Curvature perturbation in power law (e.g. extended)
 inflation,''
 Phys.\ Lett.\  B {\bf 274} (1992) 168;
 %%CITATION = PHLTA,B274,168~;%%
I.~P.~C.~Heard and D.~Wands,
  ``Cosmology with positive and negative exponential potentials,''
  Class.\ Quant.\ Grav.\  {\bf 19} (2002) 5435
  [arXiv:gr-qc/0206085];
  %%CITATION = CQGRD,19,5435;%%




\bibitem{kasner}
  E.~Kasner,
  ``Geometrical theorems on Einstein's cosmological equations,''
  Am.\ J.\ Math.\  {\bf 43} (1921) 217.
  %%CITATION = AJMAA,43,217;%%

\bibitem{singtheorems}
  S.~W.~Hawking and R.~Penrose,
  ``The Singularities of gravitational collapse and cosmology,''
  Proc.\ Roy.\ Soc.\ Lond.\ A {\bf 314} (1970) 529;
  %%CITATION = PRSLA,A314,529;%%
for a review, see e.g. A.~Borde and A.~Vilenkin,
  ``Singularities in inflationary cosmology: A Review,''
  Int.\ J.\ Mod.\ Phys.\ D {\bf 5} (1996) 813
  [gr-qc/9612036].
  %%CITATION = GR-QC/9612036;%%

\bibitem{bkl}
 V.~A.~Belinsky, I.~M.~Khalatnikov and E.~M.~Lifshitz,
  ``Oscillatory approach to a singular point in the relativistic cosmology,''
  Adv.\ Phys.\  {\bf 19} (1970) 525;
  %%CITATION = ADPHA,19,525;%%
 T.~Damour, M.~Henneaux, B.~Julia and H.~Nicolai,
  ``Hyperbolic Kac-Moody algebras and chaos in Kaluza-Klein models,''
  Phys.\ Lett.\ B {\bf 509} (2001) 323
  [hep-th/0103094];
  %%CITATION = HEP-TH/0103094;%%
  T.~Damour, M.~Henneaux and H.~Nicolai,
  ``Cosmological billiards,''
  Class.\ Quant.\ Grav.\  {\bf 20} (2003) R145
  [hep-th/0212256].
  %%CITATION = HEP-TH/0212256;%%


\bibitem{Bachas:1999um}
K.~Dasgupta, D.~P.~Jatkar and S.~Mukhi,
  ``Gravitational couplings and Z(2) orientifolds,''
  Nucl.\ Phys.\ B {\bf 523} (1998) 465
  [hep-th/9707224];
  %%CITATION = HEP-TH/9707224;%%
  C.~P.~Bachas, P.~Bain and M.~B.~Green,
  ``Curvature terms in D-brane actions and their M theory origin,''
  JHEP {\bf 9905}, 011 (1999)
  [hep-th/9903210].
  %%CITATION = HEP-TH/9903210;%%

\bibitem{dbi}
 E.~Silverstein and D.~Tong,
  ``Scalar speed limits and cosmology: Acceleration from D-cceleration,''
  Phys.\ Rev.\ D {\bf 70} (2004) 103505
  [hep-th/0310221];
  %%CITATION = HEP-TH/0310221;%%
 M.~Alishahiha, E.~Silverstein and D.~Tong,
  ``DBI in the sky,''
  Phys.\ Rev.\ D {\bf 70} (2004) 123505
  [hep-th/0404084].
  %%CITATION = HEP-TH/0404084;%%

\bibitem{higherder}
 L.~Amendola,
  ``Cosmology with nonminimal derivative couplings,''
  Phys.\ Lett.\ B {\bf 301} (1993) 175
  [gr-qc/9302010];
  C.~Germani and A.~Kehagias,
  ``New Model of Inflation with Non-minimal Derivative Coupling of Standard Model Higgs Boson to Gravity,''
  Phys.\ Rev.\ Lett.\  {\bf 105} (2010) 011302
  [arXiv:1003.2635 [hep-ph]];
  %%CITATION = ARXIV:1003.2635;%%
  %%CITATION = GR-QC/9302010;%%
  S.~Tsujikawa,
  ``Observational tests of inflation with a field derivative coupling to gravity,''
  Phys.\ Rev.\ D {\bf 85} (2012) 083518
  [arXiv:1201.5926 [astro-ph.CO]].
  %%CITATION = ARXIV:1201.5926;%%



\bibitem{gkp}
S.~B.~Giddings, S.~Kachru and J.~Polchinski,
  ``Hierarchies from fluxes in string compactifications,''
  Phys.\ Rev.\  D {\bf 66} (2002) 106006
  [arXiv:hep-th/0105097].
  %%CITATION = PHRVA,D66,106006;%%

\bibitem{duplift}
 E.~Dudas and S.~K.~Vempati,
  ``Large D-terms, hierarchical soft spectra and moduli stabilisation,''
  Nucl.\ Phys.\ B {\bf 727} (2005) 139
  [hep-th/0506172];
  %%CITATION = HEP-TH/0506172;%%
 G.~Villadoro and F.~Zwirner,
  ``De-Sitter vacua via consistent D-terms,''
  Phys.\ Rev.\ Lett.\  {\bf 95} (2005) 231602
  [hep-th/0508167];
  A.~Achucarro, B.~de Carlos, J.~A.~Casas and L.~Doplicher,
  ``De Sitter vacua from uplifting D-terms in effective supergravities from realistic strings,''
  JHEP {\bf 0606} (2006) 014
  [hep-th/0601190];
  %%CITATION = HEP-TH/0601190;%%
  %%CITATION = HEP-TH/0508167;%%
    K.~Choi and K.~S.~Jeong,
  ``Supersymmetry breaking and moduli stabilization with anomalous U(1) gauge symmetry,''
  JHEP {\bf 0608} (2006) 007
  [hep-th/0605108];
  %%CITATION = HEP-TH/0605108;%%
  E.~Dudas and Y.~Mambrini,
  ``Moduli stabilization with positive vacuum energy,''
  JHEP {\bf 0610} (2006) 044
  [hep-th/0607077].
  %%CITATION = HEP-TH/0607077;%%

\bibitem{fuplift}
O.~Lebedev, H.~P.~Nilles and M.~Ratz,
  ``De Sitter vacua from matter superpotentials,''
  Phys.\ Lett.\ B {\bf 636}, 126 (2006)
  [hep-th/0603047];
  %%CITATION = HEP-TH/0603047;%%
  M.~Gomez-Reino and C.~A.~Scrucca,
  ``Locally stable non-supersymmetric Minkowski vacua in supergravity,''
  JHEP {\bf 0605}, 015 (2006)
  [hep-th/0602246];
  %%CITATION = HEP-TH/0602246;%%
  E.~Dudas, C.~Papineau and S.~Pokorski,
  ``Moduli stabilization and uplifting with dynamically generated F-terms,''
  JHEP {\bf 0702}, 028 (2007)
  [hep-th/0610297];
  %%CITATION = HEP-TH/0610297;%%
H.~Abe, T.~Higaki, T.~Kobayashi and Y.~Omura,
  ``Moduli stabilization, F-term uplifting and soft supersymmetry breaking terms,''
  Phys.\ Rev.\ D {\bf 75}, 025019 (2007)
  [hep-th/0611024];
  %%CITATION = HEP-TH/0611024;%%
  R.~Kallosh and A.~D.~Linde,
  ``O'kklt,''
  JHEP {\bf 0702}, 002 (2007)
  [hep-th/0611183].  %%CITATION = HEP-TH/0611183;%%
  %%CITATION = HEP-PH/0607090;%%

\bibitem{dudas}
  E.~Dudas,
  ``Supersymmetry breaking in the effective Horava-Witten supergravity and quantization rules,''
  Phys.\ Lett.\ B {\bf 416} (1998) 309
  [hep-th/9709043].
  %%CITATION = HEP-TH/9709043;%%


\bibitem{oscil}
 E.~Ramirez,
  ``Low power on large scales in just enough inflation models,''
  Phys.\ Rev.\ D {\bf 85} (2012) 103517
  [arXiv:1202.0698 [astro-ph.CO]];
  %%CITATION = ARXIV:1202.0698;%%
  S.~Downes and B.~Dutta,
  ``Inflection Points and the Power Spectrum,''
  arXiv:1211.1707 [hep-th];
  %%CITATION = ARXIV:1211.1707;%%
S.~Avila, J.~Martin and D.~Steer,
  ``Superimposed Oscillations in Brane Inflation,''
  arXiv:1304.3262 [hep-th];
A.~Gruppuso, P.~Natoli, F.~Paci, F.~Finelli, D.~Molinari, A.~De Rosa and N.~Mandolesi,
  ``Low Variance at large scales of WMAP 9 year data,''
  arXiv:1304.5493 [astro-ph.CO];
 Z.~-G.~Liu, Z.~-K.~Guo and Y.~-S.~Piao,
  ``Obtaining the CMB anomalies with a bounce from the contracting phase to inflation,''
  arXiv:1304.6527 [astro-ph.CO].

\bibitem{planck1}
 P.~A.~R.~Ade {\it et al.}  [Planck Collaboration],
 ``Planck 2013 results. I. Overview of products and scientific results,''
  arXiv:1303.5062 [astro-ph.CO].
  %%CITATION = ARXIV:1303.5062;%%

\bibitem{planck2}
  P.~A.~R.~Ade {\it et al.}  [Planck Collaboration],
  ``Planck 2013 results. XVI. Cosmological parameters,''
  arXiv:1303.5076 [astro-ph.CO].

\bibitem{planck3}
  P.~A.~R.~Ade {\it et al.}  [Planck Collaboration],
  ``Planck 2013 results. XXII. Constraints on inflation,''
  arXiv:1303.5082 [astro-ph.CO].

\bibitem{planck4}
  P.~A.~R.~Ade {\it et al.}  [Planck Collaboration],
  ``Planck 2013 Results. XXIV. Constraints on primordial non-Gaussianity,''
  arXiv:1303.5084 [astro-ph.CO].
  %%CITATION = ARXIV:1303.5084;%%



\bibitem{Ijjas:2013vea}
  A.~Ijjas, P.~J.~Steinhardt and A.~Loeb,
  ``Inflationary paradigm in trouble after Planck2013,''
  arXiv:1304.2785 [astro-ph.CO];
  %%CITATION = ARXIV:1304.2785;%%
See however the recent talk of A. Linde in KITP,
http://online.kitp.ucsb.edu/online/primocosmo-c13/linde/oh/01.html.

\bibitem{olive}
  K.~A.~Olive,
  ``Inflation,''
  Phys.\ Rept.\  {\bf 190} (1990) 307.
  %%CITATION = PRPLC,190,307;%%



\bibitem{Zwiebach:1985uq}
  B.~Zwiebach,
  ``Curvature Squared Terms and String Theories,''
  Phys.\ Lett.\ B {\bf 156}, 315 (1985).
  %%CITATION = PHLTA,B156,315;%%



\end{thebibliography}
\end{document}